\documentclass[11pt]{article}

\usepackage{amsmath,amsthm, amssymb,amsfonts}
\usepackage{epsfig,cancel}
\usepackage{dsfont}

\numberwithin{equation}{section}

\newcommand{\beq}{\begin{equation}}
\newcommand{\eeq}{\end{equation}}
\newcommand{\ba}{\begin{array}}
\newcommand{\ea}{\end{array}}
\newcommand{\bea}{\begin{eqnarray}}
\newcommand{\eea}{\end{eqnarray}}
\newcommand{\bean}{\begin{eqnarray*}}
\newcommand{\eean}{\end{eqnarray*}}
\newcommand{\eref}[1]{(\ref{#1})}

\newcommand{\comment}[1]{}

\newcommand{\cO}{{\cal O}}
\newcommand{\cN}{{\cal N}}

\newcommand{\cA}{{\cal A}}
\newcommand{\cB}{{\cal B}}
\newcommand{\cC}{{\cal C}}
\newcommand{\cL}{{\cal L}}
\newcommand{\cV}{{\cal V}}

\def\cjn1{{\cA, \cC^*\otimes \wedge^j \cN^*}}
\def\bjn1{{\cA, \cB^*\otimes \wedge^j \cN^*}}
\def\vjn1{{\cA, \cV^*\otimes \wedge^j \cN^*}}
\def\cjn2{{\cA, \cC\otimes \wedge^j \cN^*}}
\def\bjn2{{\cA, \cB\otimes \wedge^j \cN^*}}
\def\vjn2{{\cA, \cV\otimes \wedge^j \cN^*}}

\def\fnote#1#2{\begingroup\def\thefootnote{#1}\footnote{#2}
     \addtocounter{footnote}{-1}\endgroup}

\newtheorem{mydef}{Definition}

\begin{document}

\vspace{1cm}

\title{{\huge \bf
The Atiyah Class and Complex Structure \\
Stabilization in Heterotic \\
Calabi-Yau Compactifications
}}

\vspace{2cm}

\author{
Lara B. Anderson${}^{1}$,
James Gray${}^{2}$,
Andre Lukas${}^{3}$,
Burt Ovrut${}^{1}$
}
\date{}
\maketitle
\begin{center} {\small ${}^1${\it Department of Physics, University of
      Pennsylvania, \\ Philadelphia, PA 19104-6395, U.S.A.}
    \\
    ${}^2${\it Arnold-Sommerfeld-Center for Theoretical Physics, \\
Department f\"ur Physik, Ludwig-Maximilians-Universit\"at M\"unchen,\\
Theresienstra\ss e 37, 80333 M\"unchen, Germany} \\ 
    ${}^3${\it Rudolf Peierls Centre for Theoretical Physics, Oxford
      University,\\
      $~~~~~$ 1 Keble Road, Oxford, OX1 3NP, U.K.}\\
    \fnote{}{andlara@physics.upenn.edu, james.gray@physik.uni-muenchen.de, lukas@physics.ox.ac.uk, ovrut@elcapitan.hep.upenn.edu} }
\end{center}

\abstract{\noindent Holomorphic gauge fields in $N=1$ supersymmetric
  heterotic compactifications can constrain the complex structure
  moduli of a Calabi-Yau manifold. In this paper, the tools necessary
  to use holomorphic bundles as a mechanism for moduli stabilization
  are systematically developed. We review the requisite deformation
  theory -- including the Atiyah class, which determines the
  deformations of the complex structure for which the gauge bundle
  becomes non-holomorphic and, hence, non-supersymmetric. In addition,
  two equivalent approaches to this mechanism of moduli stabilization
  are presented. The first is an efficient computational algorithm for
  determining the supersymmetric moduli space, while the second is an
  F-term potential in the four-dimensional theory associated with
  vector bundle holomorphy. These three methods are proven to be
  rigorously equivalent. We present explicit examples in which large
  numbers of complex structure moduli are stabilized. Finally,
  higher-order corrections to the moduli space are discussed.  }

\newpage

\tableofcontents

%
%

\section{Introduction}\label{intro}

Compactification of the $E_8 \times E_8$ heterotic string
\cite{Candelas:1985en,Green:1987mn,Witten:1996mz} and heterotic
M-theory \cite{Lukas:1997fg}--\cite{Lukas:1998hk} on Calabi-Yau
threefolds has provided a rich arena for string phenomenology
\cite{burt_a}-\cite{Blaszczyk:2011ib}. However, moduli stabilization in
such theories has remained a crucial and long-standing problem. Many
of the techniques available in Type IIB string constructions
 -- particularly the way in which fluxes are used to
stabilize moduli -- cannot be directly transferred to the heterotic
case. For example, using NS three-form flux to stabilize the complex
structure in this context makes it very difficult to stabilize the
remaining moduli fields with non-perturbative effects. In addition,
the introduction of flux  naively leads to non-K\"ahler
compactification manifolds, over which there are few techniques
available for explicitly constructing gauge bundles.

In recent work \cite{Anderson:2010mh,Anderson:2011cz}, we introduced a
new approach to moduli stabilization via a geometric observation: {\it
  the complex structure moduli of a Calabi-Yau threefold, $X$, can be
  constrained by the presence of a holomorphic vector bundle, $V
  \stackrel{\pi}{\rightarrow} X$, in a heterotic compactification}. In
particular, certain deformations of the complex structure, with all
other moduli held fixed, can lead to the gauge bundle becoming
non-holomorphic and, hence, non-supersymmetric. This is associated
with an F-term contribution to the potential energy which stabilizes
the corresponding complex structure moduli. In \cite{Anderson:2010mh},
this formalism was introduced and an explicit example of the
constraints on complex structure was presented (see
\cite{Donagi:2009ra}-\cite{Grimm:2010gk} for related work in other
contexts).

Vector bundle holomorphy is a significant new tool in heterotic moduli stabilization, since point-wise holomorphic vector bundles can perturbatively stabilize the complex structure in a supersymmetric Minkowski vacuum  without deforming the background away from a Calabi-Yau geometry. In \cite{Anderson:2011cz}, we proposed a scenario to stabilize all geometric moduli -- that is, the complex structure, K\"ahler moduli and the dilaton -- in heterotic Calabi-Yau compactifications without NS flux. This was accomplished using a hidden sector gauge bundle whose holomorphic structure fully stabilized the complex structure of the base. Combining this with other perturbative effects, such as slope-stability of the gauge bundle, and certain non-perturbative corrections to the superpotential, we were able to find AdS vacua in which all geometric moduli were stabilized.

Our approach can be viewed in another, more fundamental,
light. Instead of viewing it as a method of moduli stabilization, it
should be observed that our approach is simply the correct
identification of the local heterotic moduli space -- that is, the
identification of the only degrees of freedom which should have been
considered in the first place. More precisely, the flat directions of
a heterotic potential are actually counted by different quantities
than those used historically. In the bulk of the literature, heterotic
moduli have been taken to be
\\
\begin{table}[!h]
\begin{center}
\begin{tabular}{c|ccc}
Moduli &Complex Structure &K\"ahler &Bundle Moduli \\\hline
Cohomology &$H^1(X, TX)$&$H^1(X,TX^{\vee})$&$H^1(X, \textnormal{End}(V))$
\end{tabular}
\end{center}
\label{table_mod}
\end{table}

\noindent Here, $H^1(X,TX)=H^{2,1}(X)$ and $H^1(X, TX^{\vee})=H^{1,1}(X)$ are the familiar complex structure and K\"ahler moduli of the Calabi-Yau threefold $X$, while $H^1(X,\textnormal{End}(V))=H^1(X, V\otimes V^{\vee})$ are the allowed holomorphic fluctuations (for fixed complex structure) of the connection on a gauge bundle $V \stackrel{\pi}{\rightarrow} X$.
However, it is fundamentally wrong to identify these as the ``moduli'' of the supersymmetric vacuum space! In order to have an $N=1$ supersymmetric heterotic vacuum, the geometry must satisfy the Hermitian Yang-Mills equations \cite{Green:1987mn}
\beq\label{intro_hym}
F_{ab}=F_{{\bar a}{\bar b}}=0~~,~~g^{a{\bar b}}F_{a {\bar b}}=0~.
\eeq
It is well-known that given a background configuration satisfying \eref{intro_hym}, not all fluctuations of the forms in the above Table preserve these conditions. The true flat directions, that is, those which satisfy the Hermitian Yang-Mills equations, are generally complicated combinations (and subsets) of these fields. Previous work \cite{Lukas:1999nh}-\cite{Anderson:2010ty} investigated the constraints arising from $g^{a{\bar b}}F_{a{\bar b}}=0$, the slope-stability condition \cite{AG} on $V$, and the conditions this places on the combined K\"ahler and vector bundle moduli spaces.

In more recent work \cite{Anderson:2010mh,Anderson:2011cz}, the $F_{ab}=0$ condition of vector bundle holomorphy was explored. Specifically, we observed that the moduli whose fluctuations preserve this equation are not those listed in the above Table. Rather, they are a particular combination of the complex structure and vector bundle moduli. This combination is well-known in the deformation theory of compact complex manifolds \cite{atiyah, kodaira, kobayashi,kuranishi,donaldson_def}, and is defined by the cohomology group which counts the {\it actual fluctuations which preserve holomorphy}. This is  given by
\beq
H^1(X,{\cal Q}) \ ,
\eeq
where the bundle ${\cal Q}$ is constructed via the short exact sequence
\beq\label{atiyah_intro}
0 \to \textnormal{End}(V) \to {\cal Q} \to TX \to 0
\eeq
 introduced by Atiyah in \cite{atiyah}. The deformations $H^1(X, {\cal Q})$ measure the first-order simultaneous deformations of a bundle and its base in such a way that holomorphy is preserved.

The importance of this deformation theory for heterotic string phenomenology was the central observation of \cite{Anderson:2010mh}. However, having recognized the significance of bundle holomorphy for moduli stabilization,
it is crucial to develop it into a practical tool that can be applied to realistic vacua. Given a particular Calabi-Yau threefold, there are a number of questions one would like to answer. These include: 
\begin{enumerate}
\item How does one efficiently decide whether a given holomorphic vector bundle constrains the complex structure moduli of $X$ and, if so, how many such moduli are stabilized? 
\item Is there a simple method for selecting hidden sector vector bundles which will perturbatively fix all complex structure moduli of $X$? 
\item If such a class of hidden sector bundles could be found, how can one systematically and rapidly determine their properties and their compatibility with a realistic visible sector?
\end{enumerate}
We will answer these questions, at least in part, in this paper, and present a systematic study of vector bundle holomorphy and the associated stabilization mechanism within the context of several illustrative examples. In the process, we enhance the range of tools available to address holomorphic deformations and determine the moduli of a supersymmetric heterotic vacuum.

Specifically, we make use of several different and complementary points of view to gather information about the deformation space\ $H^1(X, {\cal Q})$. In addition to computing this space directly using techniques in deformation theory, we also present an alternative approach that is computationally much easier. We refer to a direct computation of the moduli space $H^1(X, {\cal Q})$ through its defining sequence \eref{atiyah_intro} as a ``top down" approach, since this requires one to start with a fixed initial vector bundle and analyze fluctuations away from the given background. The alternative approach follows a ``bottom up" point of view. This begins with the question: what geometric quantities (or ``support") must be available in order to build a holomorphic vector bundle using a given construction? Furthermore, how do these quantities depend on the complex structure moduli of $X$? This ``bottom up" approach is intuitively equivalent to the Atiyah computation of $H^1(X, {\cal Q})$. Within the context of a specific class of examples, we will demonstrate that it is, in fact, rigorously equivalent, and far simpler computationally. Finally, we show that  in certain classes of examples it possible to explicitly determine the F-terms in the four-dimensional theory that correspond to the $F_{ab}=0$ condition in \eref{intro_hym}.

The outline of this paper is as follows.  In the following section, we
briefly review the conditions for a supersymmetric heterotic vacuum
and the way that these conditions enter the four-dimensional
potential. In Section \ref{gen_theory}, we outline the theoretical
framework for our discussion by introducing the basic fluctuation of
the holomorphy condition, $F_{ab}=0$. In particular, we study this
fluctuation from a ten-dimensional point of view, via the fluctuation
of \eref{intro_hym}, and from a four-dimensional viewpoint, via the
Chern-Simons three-form, $\omega_3^{YM}=F\wedge A -\frac{1}{3}A\wedge
A \wedge A$, contribution to the Gukov-Vafa-Witten superpotential. In
Section \ref{full_atiyah}, we present the appropriate mathematical
framework for determining the moduli; namely, the deformation theory
of simultaneous holomorphic fluctuations of a bundle and its base. We
review the properties of the space, $\textnormal{Def}(X,V)$, of
simultaneous changes of the complex structure of $X$ and the
connection $A$ on $V$ such that the bundle remains holomorphic. To
this end, we develop the necessary mathematics of the Atiyah sequence
\eref{atiyah_intro} and the first order deformation space $H^1(X,{\cal
  Q})$.

In Section \ref{main_eg}, we explore the two complementary approaches to these geometric deformations within the context of a class of simple rank $2$ vector bundles -- an $SU(2)$ extension $0 \to \cL \to V \to \cL^{\vee} \to 0$ of a line bundle and its dual. First, we apply the deformation theory of Atiyah (that is, the ``top down" approach) to this class of examples. Next, we consider the construction of such bundles from first principles and demonstrate that the essential defining geometric ingredient, the extension class $\phi \in Ext^1(\cL^{\vee},\cL)$, can ``jump" with complex structure (this is the ``bottom up" approach described above). Finally, we analyze the four-dimensional effective field theory for this class of examples and compute the F-terms associated with the vector bundle holomorphy. In Section \ref{equivalent_approaches}, we prove that these three viewpoints are equivalent for the chosen class of rank $2$ extension bundles.

To demonstrate the effectiveness of this approach to moduli stabilization, we provide a concrete example. In Section \ref{our_eg}, we present a rank $2$ extension bundle whose holomorphy fixes $80$ out of $82$ complex structure moduli of the Calabi-Yau threefold base. In Subsection \ref{equiv_downstairs}, we combine this moduli stabilization with the effects of freely acting discrete automorphisms. These arise frequently in realistic heterotic compactifications. The same calculation is performed where, now, the threefold has been quotiented by a freely acting discrete symmetry, $\Gamma=\mathbb{Z}_2 \times \mathbb{Z}_3$. In this case, the quotient bundle over the non-simply connected threefold $X/\Gamma$ stabilizes $10$ out of the $11$ complex structure moduli.
To show the general applicability of our analysis, 
we explore two different classes of vector bundles whose holomorphy constrains the complex structure of the Calabi-Yau threefold. In particular, we present a class of $SU(3)$ bundles whose holomorphy depends on a tri-linear (Yoneda) product and a rank $2$ bundle, defined via the monad construction, which is holomorphic only when certain complex structure dependent bundle homomorphisms exist. 

In Section \ref{higher_order_sec}, we explore the structure of
possible higher-order corrections to the calculations presented in the
paper. In Subsection \ref{mod_space_bound}, a bound is derived on the
dimension of the simultaneous deformations space,
$\textnormal{dim}(\textnormal{Def}(X,V))$, to all orders in the
deformation expansion. Furthermore, in Section \ref{hidden_section} we
describe how bundles which are only holomorphic for isolated points in
complex structure moduli space can be used as part of a comprehensive
hidden sector mechanism to stabilize all geometric moduli in a
heterotic compactification. Finally, in Section \ref{conclusions_sec}
conclusions and directions for future work are presented.  In the
Appendices we present a collection of useful technical results. We
turn now to our first consideration -- the conditions for $N=1$
supersymmetry in a heterotic vacuum.

\subsection{Supersymmetric Heterotic Vacua}
One of the conditions for a solution of heterotic string theory to
preserve supersymmetry is that the variation of the ten-dimensional
gaugino under supersymmetry transformations should vanish. If the
manifold is taken to be a direct product of four-dimensional Minkowski
space and a Calabi-Yau threefold, this condition gives rise to the
so-called Hermitian Yang-Mills equations for zero slope,
\bea \label{HYM} g^{a \bar{b}} F_{a \bar{b}}=0 \;,\; F_{ab}=0 \;,\;
F_{\bar{a}\bar{b}}=0 \ .  \eea Here $F$ is the gauge field strength
associated with a connection $A$ on a vector bundle $V$, and $a$ and
$\bar{b}$ are holomorphic and anti-holomorphic indices on the
Calabi-Yau manifold.

Let us consider a specific ten-dimensional
field configuration for which equations \eqref{HYM} are
satisfied. If we now vary the K\"ahler and complex structure moduli of
the Calabi-Yau threefold, there is no guarantee that the field strength will continue to satisfy these equations. The first equation in \eqref{HYM} clearly
depends on the K\"ahler moduli through the presence of the
metric. If we change the K\"ahler class, it could be that no solution
to this equation will exist. This is associated with the $D$-term breaking of supersymmetry at so-called stability walls, and was discussed \cite{Sharpe:1998zu} and in detail in recent
work by the authors \cite{Anderson:2009nt,Anderson:2009sw,Anderson:2010tc,Anderson:2010ty}.
The last two equations in \eqref{HYM} are specified, in part, by the
definition of holomorphic and anti-holomorphic coordinates on the
compact space. This definition clearly depends on the complex
structure. For a fixed topology of the gauge
fields, specific changes in the complex structure of the
Calabi-Yau threefold may be such that these two equations no longer have a
solution. As discussed in \cite{Anderson:2010mh,Anderson:2011cz}, this is associated with the $F$-term breaking of supersymmetry by the complex structure moduli.

What happens in the effective field theory when the moduli evolve so
that the gauge fields break supersymmetry? One can see from the
dimensional reduction of the ten-dimensional effective action of the
$E_8 \times E_8$ heterotic theory that there will be a positive
definite potential in the non-supersymmetric parts of field space. To show this, consider the following three terms in the
ten-dimensional effective action,
\begin{eqnarray}
\label{spartial}
S_{\textnormal{partial}} = -\frac{1}{2 \kappa_{10}^2} \frac{\alpha'}{4}\int_{{\cal M}_{10}} \sqrt{-g} \left\{ \textnormal{tr} (F^{(1)})^2+\textnormal{tr} (F^{(2)})^2 - \textnormal{tr} R^2 \right\}~.
\end{eqnarray}
The notation here is standard \cite{Green:1987mn} with the field
strengths $F^{(1)}$ and $F^{(2)}$ being associated with the two $E_8$
factors in the gauge group. One consequence of the ten-dimensional
Bianchi Identity,
\begin{eqnarray}
  dH = -\frac{3 \alpha'}{\sqrt{2}} \left( \textnormal{tr} F^{(1)} \wedge F^{(1)} +\textnormal{tr} F^{(2)} \wedge F^{(2)} - \textnormal{tr} R \wedge R \right) \;,
\end{eqnarray}
is its integrability condition,
\begin{eqnarray}
\label{intcond}
\int_{M_{6}} \omega  \wedge \left( \textnormal{tr} \;F^{(1)} \wedge F^{(1)} + \textnormal{tr} \; F^{(2)} \wedge F^{(2)} - \textnormal{tr}\; R \wedge R \right) = 0 \;,
\end{eqnarray}
where $\omega$ is the K\"ahler form.  Using the fact that we are working,
to lowest order, with a Ricci flat metric on a manifold of $SU(3)$
holonomy, equation \eqref{intcond} can be rewritten as 
\begin{eqnarray}
  \int_{M_{10}} \sqrt{-g} \left( \textnormal{tr} (F^{(1)})^2+\textnormal{tr} (F^{(2)})^2 - \textnormal{tr} R^2  +2 \,\textnormal{tr} (F^{(1)}_{a \bar{b}} g^{a \bar{b}})^2   +2\, \textnormal{tr} (F^{(2)}_{a \bar{b}} g^{a \bar{b}})^2  \right. \\ \nonumber \left.-4 \, \textnormal{tr} (g^{a \bar{a}} g^{b \bar{b}} F^{(1)}_{ab} F^{(1)}_{\bar{a}\bar{b}})  -4 \, \textnormal{tr} (g^{a \bar{a}} g^{b \bar{b}} F^{(2)}_{ab} F^{(2)}_{\bar{a}\bar{b}})\right)= 0~.
\end{eqnarray}
Using this relation in \eqref{spartial}, we arrive at the result
\begin{eqnarray}
\label{finalintro}
S_{\textnormal{partial}}= - \frac{1}{2 \kappa_{10}^2} \alpha' \int_{{\cal M}_{10}} \sqrt{-g} \left\{ -\frac{1}{2} \textnormal{tr} (F^{(1)}_{a \bar{b}} g^{a \bar{b}})^2  - \frac{1}{2} \textnormal{tr} (F^{(2)}_{a \bar{b}} g^{a \bar{b}})^2 \right. \\ \nonumber \left. + \textnormal{tr} (g^{a \bar{a}} g^{b \bar{b}} F^{(1)}_{ab} F^{(1)}_{\bar{a}\bar{b}}) +  \textnormal{tr} (g^{a \bar{a}} g^{b \bar{b}} F^{(2)}_{ab} F^{(2)}_{\bar{a}\bar{b}}) \right\}~.
\end{eqnarray}
The terms in Eq.~\eqref{finalintro} form a part of the ten-dimensional
theory which does not contain any four-dimensional derivatives. It
therefore contributes, upon dimensional reduction, to the potential
energy of the four-dimensional theory. In the case of a supersymmetric
field configuration, the terms in the integrand of \eqref{finalintro}
vanish, these being built out of precisely the objects which
\eqref{HYM} sets to zero.  In this case, no potential is generated. At
first glance it seems difficult to determine the precise form of the
potential in \eref{finalintro} since the Calabi-Yau metric, $g$ and
the gauge field strength, $F$ are not known explicitly except in very
special examples (or via numerical methods such as
\cite{Braun:2007sn,Braun:2008jp,Douglas:2006hz,Anderson:2010ke,Anderson:2011ed}). However,
in Ref.~\cite{Anderson:2009nt,Anderson:2009sw} it was shown how this
can be achieved for the potential which results from the first two
terms in \eqref{finalintro}. It was found that this is a positive
D-term potential which described the loss of bundle supersymmetry due
to a variation of the K\"ahler moduli~\footnote{Note that the first
  two terms in \eqref{finalintro} are indeed a positive semi-definite
  contribution to the potential since $|g^{a \bar{b}} F_{a
    \bar{b}}|^2= (g^{a \bar{b}} F_{a \bar{b}})(g^{\bar{a} b} F_{
    \bar{a} b}) = -(g^{a \bar{b}}F_{a \bar{b}})^2$.}.  Further, in
\cite{Anderson:2010mh,Anderson:2011cz}, we showed that if the complex
structure moduli are varied so that the final two equations in
\eqref{HYM} no longer admit a solution, then the last two terms in
\eqref{finalintro} don't vanish and we again obtain a positive
definite contribution to the four-dimensional potential energy, now,
however, through a non-zero F-term. This effect can stabilize the
complex structure moduli of the heterotic compactification. The
purpose of this paper is to explore this latter type of supersymmetry
breaking, and the associated stabilization of complex structure, in
much greater detail.

\section{General Theory}\label{gen_theory}

In this section, we consider solutions to ten-dimensional
heterotic supergravity 
which preserve $N=1$ supersymmetry in four-dimensions. Specifically:
\begin{itemize}
\item We choose the ten-dimensional manifold to be a direct product of
  four-dimensional Minkowski space and a Calabi-Yau threefold. Taking the H-flux to vanish, this
  is a solution to the Killing spinor equation arising from the supersymmetric variation of the 
  gravitino.
\item We choose the dilaton to be constant. This is a solution
  to the Killing spinor equation arising from the supersymmetric dilatino variation.
\item We choose the gauge fields to be a connection on a poly-stable,
  holomorphic vector bundle $V$ of zero slope. Due to the theorem of
  Donaldson, Uhlenbeck and Yau \cite{duy1,duy2}, this is equivalent to fixing a solution
  to the Killing spinor equations arising from the gaugino variation
  on such a background. These equations are known as the Hermitian
  Yang-Mills equations for zero slope, and were presented in
  \eqref{HYM}. The first equation in \eqref{HYM} says
  that the gauge bundle $V$ is poly-stable with zero slope, while
  the second and third equations, being complex conjugates of one
  another, both state that $V$ is a holomorphic bundle.
\end{itemize}
In addition, we must ensure that the integrability
condition arising from the Bianchi identity \cite{Green:1987mn} for the NS two-form is
satisfied. This can be written as
\beq\label{anomaly1}
c_2(TX)-c_2(V) =[W] \ ,
\eeq
relating the second Chern classes of $V$ and the holomorphic tangent bundle $TX$. Their difference must be the class $[W] \in H_2(X,\mathbb{Z})$ of a holomorphic curve $W$, wrapped by five-branes.

Given such a background, one can discuss the moduli space of
fluctuations about it. This is the set of possible field perturbations
which continue to solve the above equations. Upon dimensional
reduction, these fluctuations correspond to flat directions in the
effective potential -- that is, massless moduli fields in the
four-dimensional theory.  In heterotic theory, these comprise
fluctuations of the dilaton, the metric (in the form of perturbations
to the K\"ahler class and the complex structure of the Calabi-Yau
threefold) and the gauge fields. From the point of view of satisfying
the Killing spinor equations arising from the gravitino variation, all
variations of the K\"ahler class and complex structure of the
Calabi-Yau threefold are allowed. Hence, they are normally all thought
of as moduli. However, this conclusion is based on an assumption --
namely, that these fluctuations are compatible with a solution to the
other variation equations - notably the gaugino variation leading to
\eqref{HYM}. In previous papers \cite{Anderson:2009nt,Anderson:2009sw}
and \cite{Anderson:2010mh}, we discussed the implications of the fact
that this is generally not true for the K\"ahler moduli and complex
structure moduli respectively. In this paper, we greatly elaborate
upon the latter of these two observations.

\subsection{10D field equations} \label{10d}

We wish to consider the structure of the solutions to \eqref{HYM} as
one varies the complex structure of the Calabi-Yau threefold. In this
context, it is clearly inconvenient to express the equations in the
holomorphic indices associated with a fixed complex structure, as was
done in \eqref{HYM}. The equations can be reformulated in terms of an
arbitrary set of real coordinates using of the ``projector''
$P_{\mu}^{\;\nu}=(\mathds{1}_{\mu}^{\; \nu} + i J_{\mu}^{\; \nu})$ and
its conjugate $\bar{P}_{\mu}^{\; \nu}=(\mathds{1}_{\mu}^{\; \nu} - i
J_{\mu}^{\; \nu})$, where $J$ is the complex structure tensor and
$\mu,\nu=1,\dots,6$. The HYM equations \eqref{HYM} can then be written
as \bea \label{HYM111} g^{\mu \nu} P_{\mu}^{\gamma}
\bar{P}_{\nu}^{\delta} F_{\gamma \delta} &=& 0 \ ,\\ \label{HYM21}
P_{\mu}^{\; \nu} P_{\rho}^{\; \sigma} F_{\nu \sigma} = 0 &,&
\bar{P}_{\mu}^{\; \nu} \bar{P}_{\rho}^{\; \sigma} F_{\nu \sigma} = 0 \
.  \eea This makes the complex structure dependence of the equations
explicit.

Let us start with a solution to \eqref{HYM111} and \eqref{HYM21} for a
specific choice of complex structure and then vary the complex
structure, keeping the K\"ahler class of the Calabi-Yau
threefold fixed. In addition we will not vary the bundle moduli. The bundle moduli correspond to
perturbations of the gauge field which are elements of $H^1(V \otimes
V^{\vee})$. They can always be added to the connection without spoiling a
solution to \eqref{HYM111} and \eqref{HYM21}. We {\it will}, however,
allow arbitrary non-harmonic variations of the gauge field to
occur as we change the complex structure. These correspond to the
possible ways in which the gauge field can adjust in an attempt to
satisfy equations \eqref{HYM111} and \eqref{HYM21} for the perturbed
complex structure. It should be noted that not varying the K\"ahler
and bundle moduli is a choice we are making for ease of
exposition. Including the effects of such changes to the ten-
dimensional solution is a trivial extension of our analysis and does
not change our results.

We will denote unperturbed objects with a superscript ``$(0)$'' and
our small changes with a $\delta$. Thus, we have the perturbation to
the complex structure, $J = J^{(0)} + \delta J$, which induces $P=
P^{(0)} + \delta P$ and the perturbed gauge connection $A = A^{(0)} +
\delta A$. Substituting these into \eqref{HYM21} gives one the
constraints on the change in complex structure for which it is
possible to vary $A$ and still have a solution to that equation --
that is, it tells us for which variations of complex structure the
connection can adjust itself so as to remain holomorphic.  It is
expedient to write the results in terms of holomorphic and
anti-holomorphic coordinates associated with the unperturbed complex
structure $J^{(0)}$.  Demanding that both $J$ and $J^{(0)}$ square to
one tells us that, in these coordinates, the only non-zero components
of the variation of $J$ are $\delta J_{a}^{\; \bar{b}}$ and $\delta
J_{\bar{a}}^{\; b}$. Demanding that both $J$ and $J^{(0)}$ have
vanishing Nijenhuis tensor (the definition of an integrable complex
structure) tells us that the perturbations to $J^{(0)}$ are harmonic,
and thus allows us to write $\delta J_{a}^{\; \bar{b}} = -i \bar{v}_{I
  a}^{\bar{b}} \delta \mathfrak{z}^I$, where $v_I$ are tangent bundle
valued harmonic one-forms and $\delta\mathfrak{z}^I$ are variations of
the complex structure moduli $\mathfrak{z}^I$. Substituting this into
\eqref{HYM21} yields, to first-order in small perturbations, the
equation \bea \label{way2} \delta \mathfrak{z}^I v_{I [\bar{a}]}^c
F^{(0)}_{|c|\bar{b}]} + 2 D^{(0)}_{[\bar{a}} \delta A_{\bar{b}]} = 0 \
.  \eea The first term in \eqref{way2} is the amount of the original
(1,1) part of the field strength that gets rotated into (0,2)
components by the change in complex structure. The second term is the
change in the, initially vanishing, (0,2) part of the field strength
due to a change in the gauge connection.

If for a given $\delta \mathfrak{z}^I$ there is a solution to
\eqref{way2} for $\delta A$, then the resulting complex structure
deformation -- in combination with the gauge field change $\delta A$
-- is a modulus. That is, the gauge field can adapt to stay
holomorphic as the complex structure of the Calabi-Yau threefold
varies. If, however, there is no solution to \eqref{way2} for a given
$\delta \mathfrak{z}^I$, then no such adaptation of the connection is
possible. It follows that the complex structure deformation can not
preserve supersymmetry and is not a modulus of the
compactification. The associated complex structure field in the
four-dimensional effective theory will then be massive -- something we
will see explicitly in Section \ref{secway3}.  In other words, any
complex structure moduli which do not obey equation \eqref{way2} for
some $\delta A$ are unambiguously stabilized by the structure of the
vector bundle $V$.

\subsubsection{Induced Fluctuations of $F^{1,1}$}
\label{nocon} 
If there exists no $\delta A$ which satisfies \eqref{way2} for a given
$\delta \mathfrak{z}^I$, then the gauge connection can not remain
holomorphic under such a change of complex structure and supersymmetry
is unambiguously broken. If, however, such $\delta A$'s do exist, one
still has not shown that the associated change in complex structure
can preserve supersymmetry. We must show that in addition to
satisfying \eqref{way2}, $\delta A$ can also solve equation \eqref{HYM111}.
Perturbing \eqref{HYM111} in the same manner as we did \eqref{HYM21},
one obtains, to linear order in the fluctuations, \bea \label{otherbit}
g^{(0) a \bar{b}} 2 D^{(0)}_{[a} \delta A_{\bar{b}]} =0 \ .\eea We
have again written the result in terms of holomorphic indices
associated with the unperturbed complex structure. We now show that
\eqref{otherbit} can always be satisfied.

For a given $\delta \mathfrak{z}^I$, the most general solution
 to \eqref{way2} can be written as $\delta A_{\bar{a}} = \delta
 \tilde{A}_{\bar{a}} + D^{(0)}_{\bar{a}} \Lambda$, where $\Lambda$ is any
 bundle valued function and $\delta \tilde A$ is any specific
 solution to \eqref{way2}.  An arbitrary harmonic bundle
 valued one-form can always be added to this expression. However, this simply
 corresponds to a shift in the bundle moduli which we have already
 declared to be fixed.
Substituting this expression for $\delta A$ into
  \eqref{otherbit}, we find the following equation for $\Lambda$, 
   \bea\label{james}
  g^{(0) a \bar{b}} \partial_{a} \partial_{\bar{b}} \Lambda + S =0 \ ,
  \eea 
  where we have defined the quantity $S=g^{(0) a \bar{b}} 2
  D^{(0)}_{[a} \delta \tilde{A}_{\bar{b}]}$.
To proceed further, one requires an elementary result in elliptic theory \cite{Aubin} that tells us
  that this equation has a solution if and only if $S$ integrates to
  zero over the Calabi-Yau threefold. This vanishing is known to hold (see \cite{kobayashi_hitchin} for a discussion).

Finally, if we impose the condition that $\delta A \to 0$ when
  $\delta \mathfrak{z}^I \to 0$, then the solution for $\Lambda$, whose
  existence has been demonstrated above, is unique.
We conclude that if a solution to \eqref{way2} exits, then there
is a unique $\delta A$ which also satisfies \eqref{HYM111} and goes to
zero as we take $\delta {\mathfrak{z}} \to 0$. Therefore, in investigating
whether complex structure moduli are stabilized by the presence of the
gauge bundle, one need only ask if there is a solution to \eqref{way2}.
In summary, to decide whether
or not the structure of the gauge bundle fixes/does not fix a complex
structure modulus, one simply has to determine whether equation
\eqref{way2} does not/does have a solution. If there exists any
solution to \eqref{way2} for a given $\delta \mathfrak{z}^I$, then one
is guaranteed that there is a solution which simultaneously satisfies
\eqref{otherbit} as well.

\subsection{The 4D Field Theory} \label{secway3} 

The structure that we have investigated in the previous sections
appears in the four-dimensional theory through dimensional
reduction. To see this, note that the background gauge configuration
determined by $V$ contributes to the three-form field strength, which
may be locally written as, \beq\label{h_def}
H=dB-\frac{3\alpha'}{\sqrt{2}}(\omega^{3YM}-\omega^{3L}) \ , \eeq
where $\omega^{3YM}$, and $\omega^{3L}$ are the gauge and
gravitational Chern-Simons forms. In particular,
$\omega^{3YM}=\textnormal{tr} (F \wedge A -\frac{1}{3}A\wedge A \wedge
A)$.  The $H$ field provides a contribution to the four-dimensional
theory via the Gukov-Vafa-Witten superpotential \cite{Gukov:1999ya}
\beq W=\int_X \Omega \wedge H \ .  \eeq This expression is a function
of the complex structure moduli $\mathfrak{z}^a$ and fields $C_i$
descending from the ten-dimensional gauge fields. These parametrize
the volume form $\Omega$ and $\omega^{3YM}$ respectively. It is
important to note that we take the $dB$ term in \eref{h_def} to be
topologically trivial globally and, hence, it does not lead to
$[H]$-flux which could deform the base geometry away from $SU(3)$
holonomy. Instead, the geometry of $X$ is a complex, K\"ahler,
Calabi-Yau threefold (with non-vanishing Ricci tensor only at order
$\alpha'$ \cite{Witten:1986kg}).

It is well-established that the holomorphic gauge field strength
$F_{ab}$ dimensionally reduces to give terms of the form
$\frac{\partial W}{\partial C_i}$ in the four-dimensional theory. In
this section, we are restricting our discussion to a supersymmetric
Minkowski vacuum. Hence, $W=0$ and the F-terms are of the form
\beq\label{gvw_fterm} F_{C_{i}}=\frac{\partial W}{\partial
  C_i}=-\frac{3\alpha'}{\sqrt{2}}\int_X \Omega \wedge \frac{\partial
  \omega^{3YM}}{\partial C_{i}} \ .  \eeq Equation \eqref{gvw_fterm}
follows from the fact that only $\omega^{3YM}$ depends on the fields
$C_i$. For any initial complex structure $\mathfrak{z}^{(0)a}$ for
which the connection $A^{(0)}$ is holomorphic and supersymmetric, all
$F_{C_i} = 0$.

In this context, let us repeat the analysis of the fluctuations
$\delta \mathfrak{z}, \delta A$. The variation $\delta \mathfrak{z}^a$
of the complex structure will induce a fluctuation in $\Omega$. For
the gauge connection \beq A_{\mu} = A^{(0)}_{\mu} + \delta
A_{\mu}+\bar{\omega}^{i}_{\mu} \delta C_{i} + \omega^{i}_{\mu}\delta{
  \bar{C}}_{i} \ , \eeq where the ${\bar \omega}$ are harmonic forms
with respect to the background connection $A^{(0)}$ and $\delta C_i$
are the variations of $C_i$. To linear order in $\delta
\mathfrak{z}^a$ and $\delta A$, the fluctuation of the F-term in
\eref{gvw_fterm} gives \beq \delta(F_{C_i}) = \int_X
\epsilon^{\bar{a}\bar{c}\bar{b}} \epsilon^{abc} \Omega^{(0)}_{abc} 2
\bar{\omega}^{x i}_{\bar{c}} \, \textnormal{tr}(T_{x}T_{y})\left(
  \delta {\mathfrak{z}}^{I} v_{I
    [{\bar{a}}}^{c}F_{|c|{\bar{b}}]}^{(0)y} + 2D^{(0)}_{[{\bar{a}}}
  \delta A^{y}_{{\bar{b}}]} \right) \ . \label{Fint} \eeq Clearly, for
complex structure deformations $\delta\mathfrak{z}^a$ for which there
exists $\delta A$ satisfying $\delta {\mathfrak{z}}^{I} v_{I
  [{\bar{a}}}^{c}F_{|c|{\bar{b}}]}^{(0)y} + 2D^{(0)}_{[{\bar{a}}}
\delta A^{y}_{{\bar{b}}]}=0$, all $F_{C_i}$ terms vanish. It follows
that these deformations are not obstructed by the potential energy
and, hence, are complex structure moduli. On the other hand, for
deformations $\delta\mathfrak{z}^a$ for which there is no $\delta A$
which sets the integrand to zero, at least one $F_{C_i}$ term is
non-vanishing. The corresponding complex structure deformations are
then obstructed by a positive potential and, hence, these fields are
massive and fixed at their initial value. The key point is that the
bracket in \eqref{Fint} is identical to the left-hand side of
Eq.~\eqref{way2} which was derived from the Hermitian Yang-Mills
equations and, hence, that the conclusions from the 10-dimensional and
the four-dimensional effective theories are consistent.

The above discussion is subject to the following caveat. The complex
structure deformations obstructed in ten-dimensions are not zero-modes
and generically have a mass of the same order as other heavy states
descending from the gauge fields. Hence, they should not really be
regarded as fields in the four-dimensional effective theory. There are
examples, however, when this mass, although non-zero, will be
suppressed relative to other mass scales \cite{Anderson:2011cz}. In
such cases, the four-dimensional discussion of fixing some complex
structure moduli is valid. We will provide examples in later sections
in which regions of moduli space can be found where the fluctuations
$\delta \mathfrak{z}^a$ are comparatively light.

\section{The Atiyah Class}\label{full_atiyah}
\subsection{The Atiyah Class and Simultaneous Deformations}
In the previous section, we considered the simultaneous variation of the complex structure and vector bundle moduli. In this section, we turn to the mathematical description of these fluctuations in terms of deformations of the Calabi-Yau threefold $X$ and a vector bundle $V$ over it. 

The relevant setting for such a discussion is mathematical deformation theory \cite{kodaira,kobayashi} and there exist powerful tools available to analyze generic fluctuations. The familiar moduli of the four-dimensional effective theory correspond to infinitesimal deformations of the complex compactification geometry which preserve the local holomorphic structure -- that is, the holomorphic structure of both $X$ and $V$. What we referred to as the complex structure and vector bundle moduli in Section \ref{gen_theory} are associated with the following objects in deformation theory. 
\begin{mydef}
Let $\textnormal{Def}(X)$ denote the space of deformations of $X$ as a complex manifold. To first-order, these deformations are parametrized by the vector space $H^1(X,TX)=H^{2,1}(X)$. These are the {\it complex structure} deformations of $X$.
\end{mydef}
\begin{mydef}
  {\it For a fixed value} of the complex structure moduli, that is,
  for a fixed complex manifold $X$, let $\textnormal{Def}(V)$ denote
  the deformation space of the vector bundle $V$. The first-order
  deformations\footnote{In fact, the cohomology that counts the gauge
    singlets arising from the vector bundle in a heterotic
    compactification is $H^1(X, \textnormal{End}_0(V))$ -- the
    $(0,1)$-forms valued in the {\it traceless endomorphims} of
    $V$. However, since $H^1(X, \textnormal{End}_0(V))=H^1(X,
    \textnormal{End}(V))$ where $\textnormal{End}(V)=V\times
    V^{\vee}$, we will frequently simply refer to the bundle moduli as
    arising from $V \otimes V^{\vee}$.} of the vector bundle are
  measured by $H^1(\textnormal{End}(V))=H^1(V\times V^{\vee})$. These
  are called the {\it bundle moduli} of $V$.
\end{mydef}
These two quantities are familiar quantities in heterotic compactifications. However, what one is ultimately interested in is the vacuum space of our theory, and this corresponds to a different object in deformation theory.
\begin{mydef}
The space of {\it simultaneous holomorphic deformations of $V$ and $X$} is denoted by $\textnormal{Def}(V,X)$. The tangent space to this first order deformation space is given by a cohomology, $H^1(X, {\cal Q})$, where ${\cal Q}$ is defined by the following short exact sequence
\beq\label{atiyah_seq}
0 \to V\otimes V^{\vee} \to {\cal Q} \stackrel{\pi}{\to} TX \to 0~.
\eeq
\end{mydef}
\noindent This sequence was first introduced in a classic paper
\cite{atiyah} by Atiyah and we will refer to it as the ``Atiyah
Sequence''. The sequence in \eref{atiyah_seq} can be derived from the
notion of a simultaneous deformation space $\textnormal{Def}(V,X)$ as
deformations of the total space of the bundle $\pi: V \to X$. A brief
review of this derivation is given in Appendix
\ref{atiyah_total_space}. For now, we simply note that, by definition,
$H^1(X,{\cal Q})$ describes the simultaneous infinitesimal
deformations of $X$ and $V$ which preserve their structure as complex
spaces and this is precisely what we must understand in heterotic
theory. To illustrate the utility of this definition of ${\cal Q}$, we
need only examine the long exact sequence in cohomology associated
with \eref{atiyah_seq}. Since $TX$ is a stable bundle
$H^0(TX)=H^3(TX)=0$ and the long exact sequence takes the form
\beq\label{atiyah_coh} 0 \to H^1(X,V\otimes V^{\vee}) \to H^1(X, {\cal
  Q}) \stackrel{d\pi}{\to} H^1(X,TX) \stackrel{\alpha}{\to}
H^2(X,V\otimes V^{\vee}) \to \ldots \eeq First, since $H^1(X,V\otimes
V^{\vee})$ injects into $H^1(X,{\cal Q})$, the familiar vector bundle
moduli are clearly a sub-space of $H^1(X,{\cal Q})$. This is in
agreement with the observation made in the section on the field theory
analysis, just under equation \eqref{HYM21}, that any change to the
bundle moduli preserves a solution to \eqref{HYM111} and
\eqref{HYM21}.

What about the complex structure moduli? The projection $d \pi$ tells
us which of the deformations of the complex structure of the base
Calabi-Yau threefold descend from the allowed deformations of the
total system. If an element in $H^1(X,TX)$ is in the image of $d \pi$,
it corresponds to a complex structure deformation which can be
obtained from the allowed deformations of the bundle $H^1(X,{\cal
  Q})$. If not, then that deformation of the complex structure is not
compatible with maintaining a holomorphic bundle. If the map $d\pi$
from $H^1(X,{\cal Q})$ to $H^1(X,TX)$ is surjective, then it follows
that the relevant cohomology group splits into two pieces \bea
H^1(X,{\cal Q})=H^1(X,V\otimes V^{\vee}) \oplus H^1(X,TX) \ .  \eea
That is, if the map $d\pi$ is surjective, then for every element of
$\textnormal{Def}(X)$ there corresponds an associated element of
$\textnormal{Def}(V,X)$. In this case, the simultaneous deformation
space maps onto the space of complex structure. In other words, for
each value of the complex structure moduli, the bundle is holomorphic.

However, in general the map $d\pi$ is {\it not surjective}! In general, all one can say is that
\beq\label{real_moduli}
H^1(X,{\cal Q})=H^1(X,V\otimes V^{\vee}) \oplus \textnormal{Im}(d\pi)
\eeq
and $\textnormal{Im}(d\pi)$ is only a subset of the complex structure moduli $H^1(X,TX)$. In this case, there exists some values of the complex structure moduli for which the bundle cannot be made holomorphic -- that is, which do not correspond to points in the simultaneous deformation space $\textnormal{Def}(V,X)$. 
It is difficult to formulate the map $d\pi$ explicitly since we have defined ${\cal Q}$ itself indirectly. However, since \eref{atiyah_coh} is exact, it follows that $\textnormal{Im}(d\pi)=\textnormal{Ker}(\alpha)$. Thus, we can determine the properties of $d\pi$ by considering the map $\alpha \in H^1(X, V \otimes V^{\vee} \otimes TX^{\vee})$. This map, called the {\it Atiyah class}, was introduced by Atiyah in \cite{atiyah} and is defined to be
\beq\label{atiyah_map_def}
\alpha=[F^{1,1}] \in H^1(V\otimes V^{\vee} \otimes TX^{\vee}) \ .
\eeq
The Atiyah class is the cohomology class of the $\{1,1\}$-component of the field strength (evaluated at a specific starting background which we are deforming away from)\footnote{Note that to define $\textnormal{Def}(V,X)$ we must first specify the background we will deform away from. That is, we must choose a valid, holomorphic starting point.}. By exactness, the actual moduli of the supersymmetric heterotic vacuum include only those elements $\nu \in H^1(X,TX)$ for which

\beq\label{good_mod}
\alpha(\nu)=0 \in H^2(X,V\otimes V^{\vee}) \ .
\eeq
The image of $\alpha$ measures the complex structure moduli that are fixed.

To make contact with the analysis of the proceeding subsections, let
us write out the condition for an element of $H^1(X,TX)$ to be in
$\textnormal{Ker}(\alpha)$ in form notation. There is a one-to-one
correspondence between elements of $H^1(X,TX)$ and tangent bundle valued
harmonic one-forms. As in earlier sections, we denote a basis of such
forms by $v_I$ and a given element in terms of the linear combination
$\delta {\mathfrak z}^I v_I$. From \eref{atiyah_map_def}, the map
$\alpha$ is simply the cohomology class of the field strength we are
perturbing. The image of an element of $H^1(X,TX)$, described above,
under the $\alpha$ map is then given by the expression
\bea 
\label{im} \delta {\mathfrak z}^I v_{I [\bar{a}}^a F^{(0)}_{|a|
  \bar{b}]} \ .
   \eea 
   This is a bundle valued $(0,2)$ form, consistent with
being an element of the target space $H^2(X,V \otimes V^{\vee})$. We hit the
zero element of the the target cohomology if and only if this image is
an exact form -- that is, if there exists some bundle valued one form
$\Gamma_{\bar{a}}$ such that 
\bea \label{way1} \delta {\mathfrak
  z}^I v_{I [\bar{a}}^a F^{(0)}_{|a| \bar{b}]} = - 2
D^{(0)}_{[\bar{a}} \Gamma_{\bar{b}]} \;.
\eea 
This condition for the
$\delta {\mathfrak z}$ to be perturbations which leave the bundle
holomorphic, is the same as those we found via the ten-dimensional
supersymmetry analysis in equation \eqref{way2}. In making this
correspondence, we have set $\Gamma_{\bar{b}} = \delta A_{\bar{b}}$.
This is permissible since both are any bundle-valued one-form
which solves the equation\footnote{Note that this description of the
  map $\alpha$ in the long exact sequence \eqref{atiyah_coh} is
  consistent with the elements of the source and target, as well as
  the map, actually being equivalence classes of forms, with two
  object being identified if they differ by an exact piece. Adding an
  exact piece to $\delta {\mathfrak z}^I v_I$ does not change the
  cohomology class of the image \eqref{im}. A similar statement holds
  for the map itself. Thus any representative elements of the
  appropriate equivalence classes for the source and map spaces may be
  chosen in performing the computation without changing the result.}.

Thus, the Atiyah class measures which deformations of the total space can preserve holomorphy of the system. That is, for which directions in complex structure moduli space $H^1(X,TX)$ it is possible to satisfy the equation
\beq
F_{ab}=F_{{\bar a}{\bar b}}=0 \ .
\eeq
The remainder of this paper will develop tools, provide examples and explore this Atiyah deformation structure in detail. Before moving on, we briefly return to the other half of the Hermitian Yang-Mills equations and the induced fluctuation of $g^{a{\bar b}}F_{a{\bar b}}$.

\subsubsection{Holomorphic Fluctuations and Slope-Stability}

In subsection \ref{nocon} we pointed out that equation \eqref{way2} is
all one needs to consider in determining which complex structure
moduli are fixed by the structure of the gauge bundle, and which are
not. The remaining equation, $g^{a{\bar b}}F_{a{\bar b}}=0$ can always be satisfied if \eqref{way2}
is. This result can also be understood from an algebraic geometry point of
view. Consider the following three-step argument. 
\begin{itemize}
\item If the perturbed complex structure is in the kernel of the map
  $\alpha$, that is, if \eqref{way2} admits a solution, then, by
  definition, the bundle is holomorphic for the perturbed complex
  structure.
\item The property of poly-stability is open in complex structure
  moduli space \cite{Huybrechts}. This means that, given that the bundle
  was taken to be polystable with respect to the initial complex
  structure, it is polystable with respect to the perturbed complex
  structure as well.
\item Finally, given the above two observations, the
  Donaldson-Uhlenbeck-Yau theorem \cite{duy1,duy2} then tells us that, for fixed
  K\"ahler and bundle moduli, there exists a unique holomorphic
  connection which also solves the equation $g^{a \bar{b}} F_{a
    \bar{b}}=0$ (equation \eqref{otherbit} in the language of subsection
  \ref{nocon}).
\end{itemize}

Thus, the discussion in ten-dimensional field theory, that is, 
differential geometry, given in Section \ref{10d} is precisely reproduced in algebraic geometry, as expected.

\section{Three Approaches to Hidden Sector Stabilization}\label{main_eg}

In previous sections, we introduced the formalism of the Atiyah class
and demonstrated that the presence of a holomorphic vector bundle over
a Calabi-Yau threefold can constrain the complex structure moduli. In
this section, we explore a concrete class of examples and develop some
of the tools necessary for computing the simultaneous deformation
space $H^1(X, {\cal Q})$. In addition, to facilitate moduli
stabilization in realistic models, we will systematically construct
classes of hidden sector gauge bundles in heterotic theories that
stabilize large numbers of complex structure moduli.

To efficiently use the tool of bundle holomorphy, we take a number of
different approaches to the problem. The first of these is the one
presented in previous sections using the deformation theory of
Atiyah. In this ``top down" approach, one specifies a heterotic
geometry (consisting of $X$ and $\pi: V \to X$) and performs the
Atiyah analysis of Section \ref{full_atiyah} to determine the
structure of the local simultaneous deformation space parametrized by
$H^1(X, {\cal Q})$. There is, however, an an alternative way to
proceed. Instead of beginning with a given geometry, a holomorphic
starting point in the total moduli space and then performing the
Atiyah computation to decide which elements of $H^1(X,TX)$ preserve
$F_{ab}=0$, one can take a ``bottom up" approach. In this case, rather
than analyze a given bundle, we would ask a different question: Given
a certain type of bundle construction, what geometric ingredients must
be available if we are to build a holomorphic bundle? Finally, in a
third approach, one could attempt to gain insight on the stabilization
of moduli from the four-dimensional effective field theory itself. As
we will see in subsequent sections, these approaches frequently encode
the same calculations in different form and, in many cases, can be
proven to be equivalent. However, they can differ widely in their
level of computational difficulty. In a given situation, information
may be more readily obtained from one method over another.

To explore this rich structure, we begin with the simplest possible vector bundles and work our way upwards in complexity. The simplest class of examples that comes to mind are line bundles. However, it is straightforward to show that in the case of line bundles, while the Atiyah class, $\alpha=[F^{1,1}]$, is generically non-vanishing, its image is always trivial. Since the target space of the Atiyah map $H^2(X,\textnormal{End}(L))=H^2(X, \cO_X)=0$ for a line bundle, the map $d\pi$ in \eref{atiyah_coh} is always surjective and a line bundle on a Calabi-Yau threefold deforms with the base $X$. Line bundles, therefore, place no constraint on the complex structure moduli.

We turn next to a slightly more complicated class of vector bundles -- an $SU(2)$ bundle defined as a non-trivial extension of a line bundle, ${\cal L}$, and its dual; that is,
\beq\label{example_bundle}
0 \to {\cal L} \to V \to {\cal L}^{\vee} \to 0 \ .
\eeq
As usual, the space of possible extensions is given by 
\beq\label{the_best_ext}
Ext^1({\cal L}^{\vee},{\cal L})=H^1({\cal L}^2)~.
\eeq
We say that an extension is ``split" or ``trivial" if $V$ is given by the direct sum $V= {\cal L} \oplus  {\cal L}^{\vee}$. The split extension corresponds to the zero element in $H^1({\cal L}^2)$, while any non-zero element of this cohomology defines an indecomposable rank $2$ bundle. 

Although we are primarily interested in the vanishing F-terms associated with the condition that $F_{ab}=0$ in the Hermitian Yang-Mills equations, we must also guarantee that any bundles we study have vanishing D-terms associated with $g^{a{\bar b}}F_{a{\bar b}}=0$ in order to have an $N=1$ supersymmetric theory. That is, we require that $V$ be a slope-stable vector bundle \cite{Green:1987mn}. In the following, we choose the line bundle $\cL$ and the K\"ahler form $\omega$ so that
\beq\label{sloped}
\mu(\cL)=\frac{1}{rank(V)}\int_X c_1(\cL)\wedge \omega \wedge \omega <0 ~.
\eeq
It is straightforward to verify that for an open region of K\"ahler moduli space, the $SU(2)$ bundle $V$ is slope-stable \cite{Anderson:2009nt} and, hence, the D-term described in Section \ref{intro} vanishes. For any such line bundle, $\cL$, it follows from \eref{sloped} that $H^0(X, \cL)=H^0(X, \cL^{\vee})=0$. Furthermore, as we derive in Appendix \ref{ext_moduli}, for fixed complex structure when the extension is non-trivial the bundle moduli are counted by 
\beq
h^1(X,V\times V^{\vee})=h^1(X,{\cal L}^2)+h^1(X,{{\cal L}^{\vee}}^2)-1 ~.
\eeq
To explicitly illustrate the behavior of this class of examples, we will further assume a simple structure for the Calabi-Yau threefold. We take $X$ to be defined as a hypersurface in a ambient space comprised of direct products of projective spaces.
As we will see in the remainder of this section, even this simple class of bundles can provide significant constraints on the complex structure moduli of $X$. We turn now to the first of our three approaches.

\subsection{Atiyah Computation}\label{section4_atiyah}
To perform the Atiyah analysis for the bundle in \eref{example_bundle}, we need to explicitly describe the source and target spaces, $H^1(X, TX)$ and $H^2(X, V\otimes V^{\vee})$ respectively, of the Atiyah map and determine {\it which element} in $H^1(V\otimes V^{\vee} \otimes TX^{\vee})$ is the Atiyah class, $\alpha$.

Let us begin with $H^1(X, TX)$, the simplest space to analyze. The tangent bundle associated with the class of CY threefolds described in this paper will be defined through a pair of short exact sequences \cite{AG}. Consider $X \subset {\cal A}$, where ${\cal A}$ is defined as the direct product of projective spaces $\mathbb{P}^{m_1} \times \ldots \times \mathbb{P}^{m_n}$. Let $X$ itself be defined by the vanishing of a polynomial $p \in H^0({\cal A}, \cN)$ for some ample line bundle $\cN$. Then the tangent bundle is described by the pair of short exact sequences 
\bea\label{tangentbundle2}
 && 0 \to \cO_{X}^{\oplus n} \stackrel{l_1}{\rightarrow} \bigoplus_{i=1,\dots,n} \cO_{X}(D_i)^{\oplus(n_i +1)} \to T{\cal A}|_{X} \to 0 \ , \\
&& 0 \to TX \to T{\cal A}|_{X} \stackrel{l_2}{\rightarrow} \cN \to 0 \ 
\eea
over $X$. The $D_i$ are the restriction of the hyperplane divisors of each projective factor of the ambient space ${\cal A}$, and the polynomial maps $l_{1,2}$ satisfy
\beq
l_2 \circ l_1 =p~ (=0~ \text{on}~X) \ .
\eeq 
It follows that the description of the complex structure moduli space $H^1(X, TX)$ is given by
\beq\label{csdefs}
H^1(X, TX)=\frac{H^0(X, \cN)}{H^0(X, T{\cal A}|_{X})} \ .
\eeq
The above quotienting of $H^0(X, \cN)$ then amounts to the degrees of freedom in the defining polynomial, $p$, modulo the $GL(k,\mathbb{C})$ transformations of the coordinates of the ambient space ${\cal A}$.

Next, we describe the target of the Atiyah map $H^2(X, V\otimes V^{\vee})$ for the bundle $V$ defined in \eref{example_bundle}. The somewhat lengthy calculation of this cohomology is carried out in Appendix \ref{ext_moduli}. Here, we simply present the final result
\beq\label{atiyahtarget}
H^2(X, V \otimes V^{\vee}) \simeq H^2(X, \cL^2) \oplus \textnormal{Ker} (\phi: H^2(X,{\cL^{\vee}}^2) \to \mathbb{C} )\ ,
\eeq
where $\phi \in H^1(X, \cL^2)$ is the extension class defining $V$ in \eref{example_bundle}.

Finally, we must describe the cohomology $H^1(X, V\otimes V^{\vee} \otimes TX^{\vee})$ containing the Atiyah class, $\alpha$. By definition, we have the following array of short exact sequences.

\beq\ba{ccccccccc}\label{display}
&&0&&0&&0&& \\
&&\downarrow&&\downarrow&&\downarrow&& \\
0&\to& {\cal L}^{\otimes 2} \otimes TX^{\vee}&\to& V\otimes {\cal L}\otimes TX^{\vee} &\to& TX^{\vee} &\to&0  \\
&& \downarrow &&\downarrow&&\downarrow&&\\
0&\to& {\cal L}\otimes V^{\vee}\otimes TX^{\vee} &\to& V\otimes V^{\vee}\otimes TX^{\vee} &\to& {\cal L}^{\vee}\otimes V^{\vee}\otimes TX^{\vee} &\to&0 \\
&&\downarrow&&\downarrow&&\downarrow&& \\
0&\to& TX^{\vee} &\to&  {\cal L}^{\vee}\otimes V \otimes TX^{\vee} & \to & {{\cal L}^{\vee}}^{\otimes 2}\otimes TX^{\vee} &\to&0 \\
&&\downarrow&&\downarrow&&\downarrow&& \\ 
&&0&&0&&0&& \\ 
\ea \eeq 

Beginning with the first column, we observe that $H^0(X, \cL^2 \otimes TX^{\vee})$ and $H^0(X, TX^{\vee})$ are both vanishing and, hence, $H^0(X, \cL \otimes V^{\vee} \otimes TX^{\vee})=0$. Moreover, $H^3(X, \cL^2 \otimes TX^{\vee})=0$ by the stability of the tangent bundle and, therefore, $H^3(X, \cL \otimes V^{\vee} \otimes TX^{\vee})=0$. The remaining cohomology of $\cL \otimes V^{\vee} \otimes TX^{\vee}$ is given by the following long exact sequence
\begin{align}\label{beta1}
&0 \to H^1(X, \cL^2 \otimes TX^{\vee}) \to H^1(X, \cL \otimes V^{\vee} \otimes TX^{\vee}) \to H^1(X, TX^{\vee}) \\\nonumber
&\stackrel{\beta_1}{\hookrightarrow} H^2(X, \cL^2 \otimes TX^{\vee}) \to  H^2(X, \cL \otimes V^{\vee} \otimes TX^{\vee}) \to  H^2(X, TX^{\vee}) \to 0  \ . \nonumber
\end{align}
We can canonically decompose the cohomology in terms of kernels and cokernels of the coboundary map, $\beta_1$, as
\beq
H^1(X, \cL \otimes V^{\vee} \otimes TX^{\vee}) \simeq H^1(X, \cL^2 \otimes TX^{\vee}) \oplus \textnormal{Ker}\left( \beta_1: H^1(X, TX^{\vee}) \to  H^2(X, \cL^2 \otimes TX^{\vee}) \right) \ .
\eeq
Similarly, the last column of \eref{display} gives us the long exact sequence
\begin{align} \label{beta2}
&0 \to H^1(X, TX^{\vee}) \to H^1(X, \cL^{\vee} \otimes V^{\vee} \otimes TX^{\vee}) \to H^1(X, {\cL^{\vee}}^2 \otimes TX^{\vee})\\\nonumber
&\stackrel{\beta_2}{\hookrightarrow} H^2(X, TX^{\vee}) \to  H^2(X, \cL^{\vee} \otimes V^{\vee} \otimes TX^{\vee}) \to  H^1(X, {\cL^{\vee}}^2 \otimes TX^{\vee}) \to 0 \nonumber
\end{align}
and, hence,
\beq
 H^1(X, \cL^{\vee} \otimes V^{\vee} \otimes TX^{\vee})\simeq H^1(X, TX^{\vee}) \oplus \textnormal{Ker}\left( \beta_2: H^1(X, {\cL^{\vee}}^{2} \otimes TX^{\vee}) \to  H^2(X,TX^{\vee}) \right) \ .
\eeq
Note that as $(0,1)$-forms both $\beta_1$ and $\beta_2$ are simply given by the extension class, $\phi \in H^1(X, \cL^2)$ associated with $V$ in \eref{example_bundle}.

Finally, we can combine this information to determine $H^1(X, V \otimes V^{\vee} \otimes TX^{\vee})$. The middle row of \eref{display} leads to the long exact sequence
\begin{align}
&0 \to H^1(X, \cL \otimes V^{\vee} \otimes TX^{\vee}) \to H^1(X,V \otimes V^{\vee} \otimes TX^{\vee}) \to H^1(X, {\cL^{\vee}} \otimes V^{\vee} \otimes TX^{\vee})\\\nonumber
&\stackrel{\beta_3}{\hookrightarrow} H^2(X, \cL \otimes V^{\vee} \otimes TX^{\vee}) \to   H^1(X,V \otimes V^{\vee} \otimes TX^{\vee}) \to  H^1(X, {\cL^{\vee}} \otimes V^{\vee} \otimes TX^{\vee})\to 0 \nonumber
\end{align}
and the cohomology of interest decomposes as
\begin{align}\label{atiyah_space}
  H^1(X,V \otimes V^{\vee} \otimes TX^{\vee}) &= H^1(X, \cL \otimes
  V^{\vee} \otimes TX^{\vee}) \\ \nonumber &\oplus \textnormal{Ker}
  \left(\beta_3: H^1(X, \cL^{\vee} \otimes V^{\vee} \otimes TX^{\vee})
    \to H^2(X, \cL \otimes V^{\vee} \otimes TX^{\vee}) \right) \\
  \nonumber
  &= H^1(X, \cL^2 \otimes TX^{\vee}) \oplus \textnormal{Ker}(\beta_1) \\
  &\oplus \textnormal{Ker} \left(\beta_3: (H^1(X, TX^{\vee}) \oplus
    \textnormal{Ker}(\beta_2)) \to H^2(X, \cL \otimes V^{\vee} \otimes
    TX^{\vee}) \right) \nonumber
\end{align}
where $\textnormal{Ker}(\beta_1)$ and $\textnormal{Ker}(\beta_2)$ are defined in \eref{beta1} and \eref{beta2} respectively.

At first view, the sub-structure of this cohomology group might seem
rather complicated. However, as we will demonstrate next, in fact only
one component of this space -- given by the simple cohomology $H^1(X,
\cL^2 \otimes TX^{\vee})$ -- is actually relevant for the Atiyah
computation for the bundle in \eref{example_bundle}.  Inspection of
\eref{atiyah_space} reveals that the cohomology group $H^1(X,V\otimes
V^{\vee}\otimes TX^{\vee})$ containing the Atiyah class
$\alpha=[F^{1,1}]$ is filtered by four components (which describe the
possible index types). These are \beq\label{pieces}
\textnormal{Ker}(\beta_1) \subset H^1(X, TX^{\vee})~~,~~H^1(X,
TX^{\vee})~~,~~H^1(X,\cL^2 \otimes
TX^{\vee})~~,~~\textnormal{Ker}(\beta_2) \subset H^1(X, {\cL^{\vee}}^2
\otimes TX^{\vee}) \ .  \eeq It is straightforward to show that only
one of these components can contain the class of the physical field
strength, $[F^{1,1}]$, associated with $V$ in
\eref{example_bundle}. Recall that for a bundle of the form $ 0 \to
\cL \to V \to \cL^{\vee} \to 0$ defined by an extension class $\phi
\in H^1(X, \cL^2)$, the transition functions take the form
\cite{Sharpe:1998zu}
\begin{eqnarray}
\label{transition_functions}
t_{ij}=\left( \ba{c c }
x_1 & \phi \\
0 &  x_2 \ 
\ea \right) \ ,
\end{eqnarray}
where $x_{1,2}$ are the transition functions associated with $\cL$ and
$\cL^{\vee}$ respectively and the lower left entry in
\eref{transition_functions} corresponds to the dual extension to
\eref{example_bundle}, parametrized by $H^1(X, {\cL^{\vee}}^2)$. For
the case at hand, this is vanishing since our background gauge
configuration associated with the bundle described by
\eref{example_bundle} corresponds to a non-trivial extension class in
$H^1(X, \cL^2)$ only. The Atiyah class can be written in terms of the
transition functions (relative to local trivializations, $t_i$ on the
patch $U_i$) as the \v{C}ech co-cycle on an intersection
$U_{ij}$. \beq \alpha=[F^{1,1}] =\{U_{ij}, t_{j}^{-1} \cdot
(t^{-1}_{ij}d t_{ij}) \cdot t_j \} \eeq As a result, the field
strength is always has the same block form as in the decomposition
given in \eref{pieces} and \eref{transition_functions}. Moreover,
since line bundles deform with their base, it is clear that the two
components of the form $H^1(X, TX^{\vee})$ in \eref{pieces} can never
contribute a non-trivial image to the Atiyah map in
\eref{atiyah_coh}. Thus, for the non-trivial extension bundle defined
in \eref{example_bundle}, the background field strength defines a the
non-zero class \beq\label{restricted_alpha} \alpha \in H^1(X, \cL^2
\otimes TX^{\vee}) \subset H^1(X,V\otimes V^{\vee}\otimes TX^{\vee}) \
.  \eeq

It follows from the special structure of the Atiyah class associated
with \eref{example_bundle} that only one component of $H^2(X, V\otimes
V^{\vee})$ in \eref{atiyahtarget} can contribute to the image of the
Atiyah map in \eref{atiyah_coh}. Because $\alpha \in H^1(X, \cL^2
\otimes TX^{\vee})$, the image of the map must be given by
\beq\label{restricted_im} H^2(X, \cL^2) \subset H^2(X, V\otimes
V^{\vee}) \ .  \eeq Note that by Serre duality \cite{AG}, this space
is dual to $H^1(X, {\cL^{2}}^{\vee})$ -- the space defining the so
called ``dual extensions" to \eref{example_bundle}.  We now have all
the ingredients we need to proceed with this ``top down''
calculation. Given an explicit bundle, one can compute the map
\beq\label{alpha_in_section4} \alpha: H^1(X,TX) \to H^2(X, \cL^2) \eeq
where $\alpha$ is defined as in \eref{restricted_alpha}. The main
point of this calculation is that, despite the complicated structure
of the space~\eqref{atiyah_space} and the source and target spaces of
the Atiyah map, we get to a relatively simple picture on how the
Atiyah map acts. Source and target spaces can be taken to be the
relatively simple cohomologies in~\eqref{restricted_im}
and~\eqref{csdefs}, and the Atiyah map itself is an element of
$H^1(X,{\cal L}^2\otimes TX^\vee)$, which is one of the components
of~\eqref{atiyah_space}. For explicit bundles, these cohomologies can
be worked out by direct computation using, for example, \v{C}ech
cohomology or, in the (multi-)projective case, the Bott-Borel-Weil
description of cohomology. We should add one final comment. It is
worth noting that, in general, given a bundle defined by extension,
and a specific extension class such as \eref{the_best_ext}, it is a
difficult problem to determine {\it which class $\alpha$} is induced
in $H^1(X,V\otimes V^{\vee}\otimes TX^{\vee})$ by the physical field
strength $[F^{1,1}]$. In later sections, we will give an example of an
explicit geometry of the form \eref{example_bundle}. In that context,
we will demonstrate that a generic extension class $\phi \in H^1(X,
\cL^2)$ induces a generic element $\alpha \in H^1(X,{\cal L}^2\otimes
TX^\vee)$.


\subsection{``Jumping" Cohomology and Extension Classes}\label{jumpingsection}
In this section, we approach the question of the moduli stabilization
induced by $V$ in \eref{example_bundle} by a second, ``bottom up"
approach. Rather than performing the Atiyah calculation described
above, we could instead ask, given a starting point in moduli space,
what are the necessary ingredients to define a holomorphic $SU(2)$
bundle of this given type? To this end, we have chosen the extension
bundle in \eref{example_bundle} not only because of its simple
structure, but also because the construction of such bundles can
depend on the complex structure in a manifest and calculable way.

To define the $SU(2)$ bundle in \eref{example_bundle}, the only
requisite ingredients are a holomorphic line bundle ${\cal L}$
satisfying $\mu(\cL) <0$, and a non-trivial extension class in $H^1(X,
\cL^2)$ which can be used to define the transition functions in
\eref{transition_functions} which locally ``glue" $\cL$ to
$\cL^{\vee}$ to form a rank $2$ non-Abelian gauge configuration.  But
what happens if we consider a line bundle $\cL$ for which
$H^1(X,\cL^2)$ is generically zero? It might seem in this case that an
indecomposable bundle of the form \eref{example_bundle} simply cannot
be defined. However, it is well known that line bundle cohomology can
``jump" over higher co-dimensional loci in complex structure moduli
space. As a result, we could begin on a locus in complex structure
moduli space for which $H^1(X, \cL^2) \neq 0$ and ask what happens to
the bundle when we perturb the complex structure away from that locus?

Consider a point $\mathfrak{z}_{0}$ on this locus and a variation
$\delta\mathfrak{z}$ to a point in complex structure moduli space for
which the Ext group $H^1(X, \cL^{2})=0$. What does the vanishing of
this Ext tell us about solutions to $F_{ab}=0$?  To answer this, first
note that, by definition, the vanishing of the extension group means
that there {\it does not exist a holomorphic, indecomposable
  $SU(2)$-valued extension bundle for the complex structure}
$\mathfrak{z}_0+\delta\mathfrak{z}$. One of the conditions
``indecomposable" or ``holomorphic" must fail. It is, in fact, the
second condition that fails, the bundle will not be holomorphic for
directions in complex structure space for which $H^1(X,
\cL^{2})=0$. The astute reader might ask: what about the direct sum
${\cal L} \oplus {\cal L}^{\vee}$? This bundle can certainly be
holomorphically defined for the new complex structure
$\mathfrak{z}_0+\delta\mathfrak{z}$. Could a deformation $\delta A$
which solves the fluctuation equation \eref{way2} break the
indecomposable bundle $V$ in \eref{example_bundle} into the split sum
of line bundles?  The answer to this question is no, for two
reasons. First, by construction, our starting background was chosen to
be in the stable region of K\"ahler moduli space. For such a choice of
K\"ahler moduli, the associated gauge connection, $A_0$, satisfying
the Hermitian-Yang-Mills equations is indecomposable and {\it cannot
  be deformed into a split connection by any infinitesimal
  transformation $\delta A$.} Even ignoring this fact, if such a
$\delta A$ existed, it is clear that since the direct sum of line
bundles, ${\cal L} \oplus {\cal L}^{\vee}$, is not poly-stable, such a
deformation $\delta A$ would break supersymmetry! Hence, the
associated $\delta\mathfrak{z}$ field would be massive and, hence,
stabilized. In fact, by the theorems stated at the end of Section
\ref{full_atiyah}, we should recall that, since {\it stability is an
  open property in complex structure moduli space}, for any pair
$\delta\mathfrak{z}$, $\delta A$ which solves \eref{way2}, there is a
solution to $g^{a\bar{b}}F_{a\bar{b}}=0$. As a result, since the
proposed $\delta \mathfrak{z}$ fluctuation would result in a $\delta
A$ violating slope-stability (that is, $g^{a\bar{b}}F_{a\bar{b}}=0$),
it is clear that {\it there exists no $\delta A$ which solves
  \eref{way2}} for a $\delta\mathfrak{z}_0$ which causes the Ext to
vanish.

In summary: when the complex structure are varied so that the Ext
group goes to zero, the bundle defined in \eref{example_bundle}, while
remaining slope-stable and indecomposable, becomes non-holomorphic. In
fact, the form defining the extension class, $\phi \in
{H^{1}}_{\mathfrak{z}_0}(X, \cL^{2})$, does not vanish when we vary
the complex structure moduli. Instead, $\phi$ is simply no longer a
closed $(0,1)$-form with respect to the new complex structure. That
is, no indecomposable extension bundle exists holomorphically.  Such
behavior should correspond to a non-trivial image for the Atiyah map
in \eref{atiyah_coh} and provides us with a second, in principle
independent, way to study the holomorphy of $V$ in
\eref{example_bundle}. Indeed, we will demonstrate in the following
sections that for simple rank $2$ extensions such as
\eref{example_bundle}, these two calculations -- the jumping Ext and
the Atiyah calculation -- give precisely the same answer.

Before we address this equivalency, let us describe how one decides
where the cohomology $H^1(X, \cL^2)$ jumps in dimension. In general,
deciding this requires a detailed knowledge of line bundle cohomology
over the threefold $X$ (and the computation can take many different
forms depending on the construction of $X$). For concreteness, we
illustrate this calculation for a simple class of CY threefolds, those
defined as a hypersurface in a product of projective spaces ${\cal
  A}=\mathbb{P}^{n_1}\times \ldots \times\mathbb{P}^{n_m}$. In this
case, $X$ is a ``favorable" manifold \cite{Anderson:2008uw} in the
sense that its Picard group is spanned by the restriction to $X$ of
ambient divisors, $D_i$, associated with the hyperplane class in
$\mathbb{P}^{n_{i}}$. A line bundle $\cL^2$ on $X$ is related to line
bundles ${\cal L}_{\cal A}^2$ on the ambient space ${\cal A}$ via the
Koszul sequence \cite{AG} \beq\label{koszully} 0 \to \cN^{\vee}
\otimes \cL_{{\cal A}}^2 \stackrel{p_{0}}{\rightarrow} \cL_{{\cal
    A}}^2 \to \cL^2 \to 0 \ , \eeq where $\cN$ is the normal bundle of
$X \subset {\cal A}$ and $p_{0} \in H^0({\cal A}, \cN)$ is polynomial
whose vanishing in ${\cal A}$ defines the hypersurface $X$. The
cohomology $H^{j}(X, \cL^2)$ is determined from the ambient space and
the defining polynomial, $p_0$, via the long exact sequence in
cohomology associated with \eref{koszully}. By the topology choice
that $\mu(\cL) < 0$ somewhere in the K\"ahler cone, it follows
\cite{AG} that $H^0(X, \cL^2)=0$ for all regions of complex structure
moduli space. In addition, for any line bundle $\cL_{{\cal A}}^2$, by
the Bott-Borel-Weil formula there is at most one non-vanishing
cohomology on ${\cal A}$. In order to have a jumping cohomology we
need this one non-vanishing cohomology to be the second one for both
${\cal N}^\vee\otimes{\cal L}_{\cal A}^2$ and ${\cal L}_{\cal
  A}^2$. Hence, the long exact sequence associated to \eqref{koszully}
takes the form \beq\label{koszul_long_exact} 0 \to H^1(X, \cL^2) \to
H^2({\cal A}, \cN^{\vee}\otimes \cL_{{\cal A}}^2)
\stackrel{p_{0}}{\rightarrow} H^2({\cal A}, \cL_{{\cal A}}^2) \to
H^2(X, \cL^2) \to 0 \ .  \eeq From \eref{csdefs}, recall that the
coefficients of the defining polynomial $p_0$ are a redundant basis
for complex structure moduli space. As these coefficients are varied
so that $p \to p_0 +\delta p$, the dimensions of
$\textnormal{Ker}(p)=h^1(X, \cL^2)$ and $\textnormal{Coker}(p)=h^2(X,
\cL^2)$ can change together, or ``jump", in such a way that the index,
$\textnormal{Ind}(\cL^2)=-h^1(X, \cL^2)+h^2(X, \cL^2)$, is preserved.

Let us consider the specific case at hand of the extension class in $H^1(X, \cL^2)$ defining the $SU(2)$ bundle in \eref{example_bundle}. If $H^1(X, \cL^2)=0$ for generic values of the complex structure, how do we decide where in complex structure moduli space it is non-vanishing? To do this we have a simple fluctuation analysis to perform.
Begin the analysis at a point $p_0$ for which $\textnormal{Ker}(p_0)=H^1(X, \cL^2) \neq 0$. We can now ask locally, which fluctuations $p=p_0 + \delta p$ satisfy $\textnormal{Ker}(p) \neq 0$? From \eref{koszul_long_exact} it is clear that we want to consider the fluctuation equation
\beq\label{jump_fluc}
(p_0 +\delta p)(k_0 + \delta k)=0 \ ,
\eeq
where $k_0 \in \textnormal{Ker}(p_0)$ and $\delta k \in H^2({\cal A}, \cN^{\vee}\otimes \cL_{{\cal A}})$. This is the condition that $\textnormal{Ker}(p) \neq 0$ for some elements $k=(k_0 + \delta k) \in H^2({\cal A}, \cN^{\vee}\otimes \cL_{{\cal A}})$.

In general, this ``jumping'' calculation for line bundle cohomology is
considerably simpler than the Atiyah computation described in the
previous section. Most notably, we have only to choose the starting
point $(k_0,p_0)$, rather than inducing the highly non-linear Atiyah
map in \eref{atiyah_map_def} and \eref{transition_functions}. Using
tools in computation algebraic geometry \cite{Gray:2008zs}, the set of
$\delta p$ solving \eref{jump_fluc} for some $\delta k$ can be readily
determined. In forthcoming work \cite{us_to_appear}, we present a
detailed mathematical and computational toolkit for analyzing such
fluctuations.

\subsection{Effective Field Theory}\label{main_fieldtheory}
As discussed in Section \ref{secway3}, in general the stabilization of
moduli induced by bundle holomorphy occurs at the compactification
scale and it is not always possible to describe this behavior in terms
of supersymmetry breaking in the $4d$ effective theory. For the class
of bundles in \eref{example_bundle}, there is a clear criteria for
when such an effective field theory description exists.

For a non-vanishing extension class {\it far} from zero, there is
generally no F-term description of the moduli stabilization associated
with the Atiyah sequence (although it should be noted that
none-the-less these degrees of freedom are completely removed from the
four-dimensional theory and their stabilized values are fully
computable). In order to discuss the F-term structure we must consider
bundles near $0 \in {\rm Ext}^{1}({\cal{L}}^{*},{\cal{L}})$.  Here,
the bundle splits as $V \to \cL \oplus \cL^{\vee}$ and its structure
group changes from $SU(2)$ to $S[U(1)\times U(1)] \simeq U(1)$. This
$U(1)$ symmetry is self commuting within $E_8$ and, as shown in
\cite{Sharpe:1998zu,Anderson:2009nt,Anderson:2009sw}, the low-energy
gauge group is enhanced by an anomalous $U(1)$ factor to $E_7 \times
U(1)$. At the hyperplane in K\"ahler moduli space where this split
bundle is polystable (supersymmetric), the bundle moduli are counted
by $h^1(X,{\cal L}^2) +h^1(X,{\cL^{\vee}}^{2})$ and become charged
under the enhanced $U(1)$ symmetry. We will denote these massless
fields by $C_{+}^{i}$ and $C_{-}^{j}$ respectively, with the subscript
$\pm$ indicating the $U(1)$ charge. The $E_7$-charged fields also
carry $U(1)$ charge, but in the following analysis we will set all
such fields to zero in the vacuum and they will not play a role in the
subsequent discussion.

Associated with the anomalous $U(1)$ symmetry is a K\"ahler
moduli dependent D-term, whose form is well-known
\cite{Sharpe:1998zu}-\cite{Anderson:2009sw}. This
four-dimensional D-term is the low energy manifestation of the
requirement that the vector bundle be poly-stable with zero
slope. Here, we simply present the D-term, using the notation of
\cite{Anderson:2009nt,Anderson:2009sw}. It is
\bea \label{mrdterm} D^{U(1)} = f - \sum_{L \bar{M}} Q^L G_{L \bar{M}} 
C^L \bar{C}^{\bar{M}}  \ ,\eea
where $C^{L}$ are the zero-mode fields with charge $Q^{L}$
under the $U(1)$ symmetry, $G_{LM}$ is a K\"ahler metric with
positive-definite eigenvalues and
\bea f =\frac{3}{16} \frac{\epsilon_S \epsilon_R^2}{\kappa_4^2}
\frac{\mu({\cal L})}{{\cal V}}\label{fiterm}
 \eea
  is a K\"ahler moduli dependent Fayet-Iliopoulos (FI) term \cite{Dine:1987xk,Sharpe:1998zu,Lukas:1999nh,Blumenhagen:2005ga,Blumenhagen:2006ux,Anderson:2009nt,Anderson:2009sw}. The
quantities
\bea \label{defstuff} \mu ({\cal L}) = d_{ijk} c_1^i({\cal L}) t^j t^k , \qquad  {\cal V} =
\frac{1}{6} d_{ijk} t^i t^j t^k \eea
are the slope of the associated line bundle ${\cal{L}}$ and the
Calabi-Yau volume respectively. Here $t^{i}$ are the K\"ahler moduli,
relative to a basis of harmonic $(1,1)$ forms $\omega_i$, with
the associated K\"ahler form given by $\omega=t^i\omega_i$. The quantities
$d_{ijk}=\int_X\omega_i\wedge\omega_j\wedge\omega_k$ are the triple
intersection numbers of the three-fold. 
The parameters $\epsilon_S$ and $\epsilon_R$ are given by
 \bea\label{epsilons}
\epsilon_S = \left( \frac{\kappa_{11}}{4 \pi} \right)^{2/3} \frac{2
  \pi \rho}{v^{2/3}}\;,\quad\epsilon_R = \frac{v^{1/6}}{\pi
  \rho} \;. \eea 
Here $v$ is the coordinate volume of the Calabi-Yau three-fold, $\rho$
is the coordinate length of the M-theory orbifold and $\kappa_{11}$ is
the eleven-dimensional gravitational constant. The four-dimensional
gravitational constant $\kappa_4$ can be expressed of these
11-dimensional quantities as $\kappa_4^2=\kappa_{11}^2/(2\pi\rho
v)$. In the subsequent discussion we will set $\kappa_{11}=1$ and
further, in order to simplify the FI term~\eqref{fiterm}, choose the
coordinate parameters $\rho$ and $v$ such that
\begin{equation}
\frac{3}{16} \frac{\epsilon_S
  \epsilon_R^2}{\kappa_4^2} =\frac{3 \pi \epsilon_S^2
  \epsilon_R^2}{16 \kappa_4^2} =1  \ .
\label{hope2}
\end{equation}

With the D-term and $U(1)$ charges in hand, it remains only to consider the superpotential. To lowest order, the four-dimensional superpotential is
 \bea \label{mrfterm} W =
\lambda_{ij}(\mathfrak{z}) C^i_+ C^j_- \ . 
\label{next2}
\eea 
%
The dimension one coefficients $\lambda_{ij}(\mathfrak{z})$ are
functions of the complex structure moduli $\mathfrak{z}^a$.  We note
that the coefficient $\lambda$ is a function of the complex structure
moduli since in fact {\it any contribution to the superpotential
  behaves linearly under rescalings, $\Delta$, of the homogeneous
  coordinates of complex structure space}. One way to see this is by
an examination of the K\"ahler potential and the scalar potential of
the four-dimensional theory. The exponential of the standard form of
the K\"ahler potential
\cite{Castellani:1990tp,Castellani:1990zd,Strominger:1990pd} scales as
$|\Delta|^2$. For the scalar potential to be invariant under
rescalings the superpotential must then scale linearly, as
stated.\footnote{Note that both superpotential and Kahler potential
  can be written in terms of the redundant, homogeneous coordinates,
  instead of the more standard affine coordinates on complex structure
  space. When one does this it is found that the overall scaling
  parameter drops out of the Lagrangian.} We discuss the scaling of
complex structure moduli in detail in Appendix \ref{hom_scaling}.

The associated F-terms are
\begin{eqnarray} \label{f1}
  F_{C^{i}_{+}}=\lambda_{ij}C_{-}^{j} + K_{C^{i}_{+}}W~&,& \quad F_{C^{j}_{-}}=\lambda_{ij}C_{+}^{i} + K_{C^{j}_{-}} W, \\ \nonumber F_{\mathfrak{z}_{\parallel}^{a}}=\frac{\partial\lambda_{ij}}{\partial{\mathfrak{z}}_{\parallel}^a}C^i_+ C^j_- + K_{{\mathfrak z}_{\parallel}^{a}} W ~&,& \quad F_{\mathfrak{z}_{\perp}^{a}}=\frac{\partial\lambda_{ij}}{\partial{\mathfrak{z}}_{\perp}^a}C^i_+ C^j_- + K_{\mathfrak{z}_{\perp}^{a}} W
\end{eqnarray}
where we have distinguished between derivatives within the sub-locus where $H^1(X, \cL^{2})\neq 0$ (specified by the coordinates $\mathfrak{z}^{a}_{\parallel}$) and those leaving this sub-locus (specified by coordinates $\mathfrak{z}^{a}_{\perp}$). Since the fields
$C_{+}^{i}$ and $C_{-}^{j}$ are zero-modes, for ${\mathfrak{z}}_{0}^{a}$ {\it on the sub-locus}, it follows that
\begin{equation}
\lambda_{ij}(\mathfrak{z}_0) =0 \quad \Rightarrow \quad \frac{\partial
 \lambda_{ij}(\mathfrak{z}_0)}{\partial {\mathfrak{z}^{a}_{\parallel}}}=0 \ .
 \label{f2}
 \end{equation}

 For the vacuum configuration associated with the $SU(2)$ bundle in \eref{example_bundle}, we will see how the $\mathfrak{z}_{\perp}^{a}$-dependence in the superpotential can
 stabilize the complex structure moduli to the sub-locus where
 holomorphic, {\it indecomposable} $SU(2)$ bundles exist. In
 performing this analysis we will look for supersymmetric Minkowski
 vacua for which $W$, as well as the F-terms \eqref{f1},
 vanishes. Given this we will not need to know the exact form of the
 K\"ahler potential in \eqref{f1}.

 First, we choose the complex structure moduli $\mathfrak{z}^{a}_0$ to
 be in the sub-locus for which the cohomologies $H^1(X, \cL^2)$ and
 $H^1(X, {\cL^{\vee}}^2)$ are non-vanishing.  This is the locus for
 which \eref{f2} holds. Note that, in this case, the superpotential
 \eqref{next2} and the first {\it three} F-terms in \eqref{f1} always
 vanish.  Now consider a bundle $V$ as in \eref{example_bundle}. This
 is defined by a non-vanishing class in ${\rm
   Ext}^{1}({\cL}^{\vee},\cL)$ and, hence, corresponds to a vacuum
 with $\langle {C_+}^{i} \rangle \neq 0$. We choose $\langle C^{i}_{+}
 \rangle$ so as to make the D-term vanish in the region of moduli
 space for which $\mu(\cL)$ and, hence, the FI term in \eref{mrdterm}
 are negative.  This one D-term constraint fixes one linear
 combination of the $C^{i}_{+}$ fields which, without loss of
 generality, can be chosen to be $\langle C_{+}^{1} \rangle$. We will
 make this choice below when it is convenient to do so.

Returning to the F-terms, this vev choice implies that the fourth F-term, $F_{\mathfrak{z}^{a}_{\perp}}$,  
in \eqref{f1} gives rise to the potential
\begin{equation}
V = |F_{\mathfrak{z}^{a}_{\perp}}|^{2}= |\frac{\partial
 \lambda_{ij}(\mathfrak{z}_0)}{\partial {\mathfrak{z}^{a}_{\perp}}} \langle C^i_+ \rangle|^{2}| C^j_-|^{2}+\dots \ ,
\label{box1}
\end{equation} 
where we suppress the multiplicative factor of $e^{K}G^{a {\bar{a}}}$
for simplicity. In contrast to Eq.~\eqref{f2}, $\frac{\partial
  \lambda_{ij}(\mathfrak{z}_0)}{\partial {\mathfrak{z}^{a}_{\perp}}}$
does not necessarily vanish. Since the function
$\lambda(\mathfrak{z})$ is non-trivial and $\lambda(\mathfrak{z}_0)=0$
while $\lambda(\mathfrak{z}_0 + \delta \mathfrak{z}_{\perp}) \neq 0$,
it is clear that $\frac{\partial^{k}
  \lambda_{ij}(\mathfrak{z}_0)}{(\partial
  {\mathfrak{z}^{a}_{\perp}})^{k}}$ will be non-trivial for some $k$. If
the first derivative in \eref{box1} is non-vanishing, one immediate
implication is that
\begin{equation}
  \langle F_{\mathfrak{z}^{a}_{\perp}} \rangle= \frac{\partial
    \lambda_{ij}(\mathfrak{z}_0)}{\partial {\mathfrak{z}^{a}_{\perp}}} \langle C^i_+ \rangle \langle C^j_-\rangle
  =0 \quad \Rightarrow \quad  <C^j_->=0   \ .
\label{hope3}
\end{equation}
Here we have assumed that all of the $C_-$ fields obtain a mass term
in this manner for simplicity. More generally, the background we are
interested in is the one where the extension class corresponding to
the $C_-$ fields is not turned on - and thus we should take
$<C^j_->=0$ anyway.  More interestingly, now consider the potential
energy obtained from all four F-terms in \eqref{f1} evaluated at a
generic point $\mathfrak{z}_0^{a}+\delta \mathfrak{z}_{\perp}^{a}$
{\it not} on the sub-locus where non-decomposable bundles $V$
exist. Then, to quadratic order in the field fluctuations we find, in
addition to the $C_{-}^{j}$ term in \eqref{box1}, that
\begin{equation}
  V = |\frac{\partial
    \lambda_{ij}(\mathfrak{z}_0)}{\partial {\mathfrak{z}^{a}_{\perp}}} \langle C^i_+ \rangle|^{2}|\delta \mathfrak{z}_\perp^{a}|^{2}+\dots \ .
\label{box4}
\end{equation}
where a sum over index $j$ is implied. This arises from the F-term, $F_{C_{-}^{j}}$, in \eqref{f1}. It follows from \eqref{box4} that any of the fluctuations in the complex structure away from the special sub-locus has a positive mass and, hence, 
\begin{equation}
\langle \delta \mathfrak{z}^{a}_{\perp} \rangle = 0 \ .
\label{box5}
\end{equation}
That is, the complex structure moduli are fixed to be on the sub-locus
where an indecomposable bundle $V$ can be holomorphic. As a final
comment, we note that it may be the case that $\frac{\partial
  \lambda_{ij}(\mathfrak{z}_0)}{\partial {\mathfrak{z}^{a}_{\perp}}}$
vanishes, but instead, for example, $\frac{\partial^2
  \lambda_{ij}(\mathfrak{z}_0)}{\partial^2
  {\mathfrak{z}^{a}_{\perp}}}\neq 0$. In this case, the stabilization
would still go through exactly as before, but the $C_{-}$ fields would
be massless to leading order and the stabilization in \eref{box4} would
occur at higher-order.

Once again, it should be noted that the above example is somewhat special in that it is
possible to give a four-dimensional description of the stabilization
of the complex structure. In general, for the mechanism presented 
here, this stabilization will take place
at high scale \cite{Anderson:2010mh}. Hence, the fixed complex structure should never have
been included as fields in the four-dimensional theory in the first
place. In such cases, one should simply write down the low-energy
$N=1$ theory without these fields present{\footnote{Indeed,
    this will even be the case in the above example if the mass term
    in equation \eqref{box4} is of the order of the compactification
    scale.}}.


\section{Equivalence of the Approaches}\label{equivalent_approaches}
In this section, we demonstrate that the three seemingly different approaches presented in Section \ref{main_eg}, namely
\begin{enumerate}
\item the Atiyah computation
\item the Jumping Extension ($H^1(X, \cL^2)$) computation
\item the four-dimensional Effective Field Theory
\end{enumerate}
are, in fact, equivalent. The correspondence between the first two approaches is highly non-trivial and requires a detailed understanding of the complex structure dependence of the holomorphic bundle $V$ in \eref{example_bundle}. Of the three possible approaches, we present the second analysis in the list as the most efficient for understanding the complex structure stabilization for this class of geometries.

\subsection{``Jumping" Ext and Atiyah Equivalence}

At first  inspection, the calculations in Sections \ref{section4_atiyah} and \ref{jumpingsection} appear very different in structure. In the Atiyah computation, as we saw in \eref{alpha_in_section4}, one must compute
\beq\label{alpha_starter}
\alpha: H^1(X,TX) \to H^2(X, V\otimes V^{\vee})~~~\text{where}~~~\alpha \in H^1(X, V\otimes V^{\vee} \otimes TX^{\vee}) \ .
\eeq
For the class of bundles in \eref{example_bundle}, we demonstrated that this simplifies to
\beq\label{good_alpha}
\alpha: H^1(X,TX) \to H^2(X, \cL^2)~~~\text{where}~~~\alpha \in H^1(X, \cL^2 \otimes TX^{\vee}) \ .
\eeq

On the other hand, we can consider the jumping of cohomology containing the extension class, $Ext^1(\cL^{\vee},\cL)=H^1(X, \cL^2)$ associated with the bundle $V$ in \eref{example_bundle}. For the Calabi-Yau threefolds considered in this paper, this cohomology is defined by the Koszul sequence
\beq\label{koszul_with_feeling}
0 \to H^1(X, \cL^2) \to H^2(\cA,\cN^{\vee}\otimes \cL^2) \stackrel{p_0}{\rightarrow} H^2(\cA, \cL^2) \to H^2(X, \cL^2) \to 0 \ .
\eeq
From this, the ``jumping" cohomology calculation in \eref{koszul_long_exact} arises from the structure of the map
\beq\label{koz}
p_0: H^2(\cA, \cN^{\vee}\otimes L^2) \to H^2(\cA, \cL^2)~~~\text{where}~~~p_0 \in H^0(\cA, \cN)~.
\eeq
To analyze where the cohomology is non-vanishing, one must solve the fluctuation equation in \eref{jump_fluc},
\beq\label{fluc_again}
(p_0 +\delta p)(k_0 +\delta k)=0
\eeq
where $k_0p_0=0$, $\delta k \in H^2(\cA, \cN^{\vee} \otimes \cL^2)$ and $\delta p \in H^0(\cA, \cN)$. The allowed complex structure moduli that keep $H^1(X, \cL^2) \neq 0$ consist of the set of $\delta p$'s which solve \eref{fluc_again} for some $\delta k$. While it is intuitively clear that the jumping cohomology calculation should give a lower bound on the Atiyah image, from the above structure, it is not obvious that these calculations are equivalent. Below, we show that this is, in fact, the case.

To begin, consider the jumping Ext calculation. In Section \ref{jumpingsection}, we showed that the Koszul sequence gives rise to jumping structure as the Calabi-Yau threefold, defined as the $p_0=0$ polynomial hypersurface, was varied. However, it is well known that not all choices of coefficients in $p_0$ give rise to different complex structures on $X$. In \eref{csdefs}, it was demonstrated that for the class of CY threefolds considered here
\beq
H^1(X,TX)=\frac{H^0(X, \cN)}{H^0(X,T\cA|_{X})}
\eeq
The quotienting removes the degrees of freedom corresponding to redefinitions of the coordinates of $\cA$. This consists of the freedom of $GL(n,\mathbb{C})$ coordinate transformations. 

Since these coordinate redefinitions are not part of the complex structure moduli space, it is clear that they cannot effect the line bundle cohomology $H^1(X, \cL^2)$ on $X$. Any such degrees of freedom must trivially drop out of the jumping calculation in \eref{fluc_again} (that is, no unphysical degrees of freedom $\delta p$ can be fixed by the constraint given in \eref{fluc_again}). As a result, we can consider the relevant $\delta p$ in \eref{fluc_again} to be $\delta p \in H^1(X,TX)$, the actual complex structure deformations around the Calabi-Yau threefold defined by $p_0=0$. With this in mind, let us examine \eref{fluc_again} once again.

Re-writing \eref{fluc_again} as
\beq\label{rewrite}
k_0\delta p = -p_0 \delta k \ ,
\eeq
consider the right-hand side of this expression. By definition, $p_0 \delta k$ is in the image of $p_0: H^2(\cA, \cN^{\vee} \otimes \cL^2) \to H^2(\cA, \cL^2)$ in \eref{koz}. Hence, this image forms the zero element of the space
\beq
H^2(X, \cL^2) = \textnormal{Coker}(p_0)
\eeq
from \eref{koz}. With this result, and the observation about $\delta p \in H^1(X, TX)$ above, \eref{fluc_again} and \eref{rewrite} may be interpreted as conditions over $X$ itself, instead of over the ambient space $\cA$. From this point of view, solving the equation in \eref{fluc_again} is equivalent to solving for the complex structure fluctuations, $\delta p$, as elements of a kernel
\beq
\delta p \in \textnormal{Ker}(k_0)
\eeq
where now
\beq\label{rethink_map}
k_0: H^1(X,TX) \to H^2(X,\cL^2) \ .
\eeq
This now looks more familiar. By inspection, the source and target of this re-writing of the fluctuation equation in \eref{fluc_again} are now exactly the source and target of the Atiyah map! What about the ``map" $k_0$ in \eref{rethink_map}?
By naive inspection of \eref{fluc_again}, we learn only that $k_0 \in H^2(\cA, \cN^{\vee} \otimes \cL^2)$ is an element of $\textnormal{Ker}(p_0)=H^1(X, \cL^2)$ since, by construction, we demand that $p_0k_0=0$. However, it is clear that when viewed as a condition over $X$ as in \eref{rethink_map}, this is not the right way to interpret the map. Among other things, the index structure of $k_0 \in H^1(X, \cL^2)$ is incorrect as a map from $H^1(X, TX)$ to $H^2(X, \cL^2)$ over $X$.

For this last piece to the puzzle, one must consider the definition of $\alpha$ in \eref{good_alpha}, recalling that $\alpha \in H^1(X, \cL^2 \otimes TX^{\vee})$. In \eref{tangentbundle2}, the holomorphic tangent bundle $TX$ was defined for the class of examples considered here. Dualizing these sequences we find,
\bea\label{tangentbundle3}
&&0 \to  T{\cal A}|_{X}^{\vee}  \stackrel{l_1^T}{\rightarrow} \bigoplus_i \cO_{X}(-D_i)^{\oplus (n_i +1)} \to \cO_{X}^{\oplus n}\to 0 \\
&&0 \to  \cN^{\vee} \to T{\cal A}|_{X}^{\vee} \stackrel{l_2^T}{\rightarrow} TX^{\vee} \to 0
\eea
with $l_1 \circ l_2 =p_0$. From this, and and the form of the line bundle $\cL$ in \eref{example_bundle}, it follows that the cohomology of interest to us is 
\begin{align}
&H^1(X, \cL^2 \otimes TX^{\vee})=\textnormal{Ker}(l_2^T) \\&l_2^T: H^2(X, \cN^{\vee} \otimes \cL^2) \to H^2(X, {T\cA}^{\vee} \otimes \cL^2)
\end{align}
By definition $\textnormal{Ker}(l_2^T) \subset H^2(X, \cN^{\vee}
\otimes \cL^2)$. However, fortuitously, by the Koszul sequence, \beq
H^2(X, \cN^{\vee} \otimes \cL^2) \subset H^2(\cA, \cN^{\vee} \otimes
\cL^2) \ , \eeq which is the same as the space containing $k_0$ in
\eref{koszul_with_feeling}. Now consider an element, $x
\in H^2(\cA, \cN^{\vee} \otimes \cL^2)$. If $x$ is an element of
$\textnormal{Ker}(l_2^T)$, then \beq l_2^T(x)=0 \ .  \eeq However,
since $p_0=l_1^T \circ l_2^T$, for any such $x$, it follows that \beq
p_0(x)=l_1^T l_2^T (x)=l_1^T(l_2^T(x))=0 \ .  \eeq Hence, $x \in
\textnormal{Ker}(p_0)$. We conclude that any map $\alpha \in H^1(X,
\cL^2 \otimes TX^{\vee})$ will play the role of $k_0$ in
\eref{rethink_map}.

Phrased more precisely, for this class of examples we have a fibration 
\beq\label{gamma_def}
\gamma: H^1(X, \cL^2) \to H^1(X, \cL^2 \otimes TX^{\vee})~
\eeq
of the space $H^1(X, \cL^2 \otimes TX^{\vee})$ containing $\alpha$ by the space $Ext^{1}=H^1(X, \cL^2)$ encoding any ``jumping''.
From the arguments above, $h^1(X, \cL^2) \geq h^1(X, \cL^2 \otimes TX^{\vee})$. Furthermore, by the definition of the extension class in the transition functions of \eref{transition_functions} and the Atiyah class, it is clear that this fibration has a zero-dimensional fiber over the origin. That is, when we are at the zero element of $H^1(X,\cL^2 \otimes TX^{\vee})$ the extension class is zero. As a result, the non-linear mapping\footnote{It is clear that $\gamma$ in \eref{gamma_def} is non-linear since $H^0(X, TX^{\vee})=0$ and, hence, there can be no linear maps from $H^1(X, \cL^2)$ to $H^1(X, \cL^2 \otimes TX^{\vee})$. This corresponds with the non-linear definition of the $\alpha=[F^{1,1}] \in H^1(X, \cL^2 \otimes TX^{\vee})$ in terms of transition functions (built from the extension class in $H^1(X, \cL^2)$) in \eref{the_best_ext} and \eref{transition_functions}. } given by the fibration $\gamma$ is surjective. Hence, if we perform the fluctuation calculation for generic values of $k_0 \in H^1(X, \cL^2)$, this corresponds to an Atiyah calculation with a generic value of $\alpha \in H^1(X, \cL^2 \otimes TX^{\vee})$.

As a result, we have shown that for a generic starting value $k_0$ of the extension class, $H^1(X, \cL^2)$, {\it the fluctuation analysis in \eref{fluc_again} is explicitly the same calculation that would be undertaken by the Atiyah mapping in \eref{alpha_starter}}. For a given example, this correspondence is even more striking when explicit polynomial descriptions of the relevant cohomology groups are used. In this paper, we have employed the Bott-Borel-Weil formalism to explicitly represent the cohomology groups as polynomial spaces (see \cite{Anderson:2008ex} for a review). In this form, it is clear in any given example that the polynomial multiplication calculations -- that is, the Groebner basis calculations \cite{Gray:2008zs} --  that must be done in \eref{alpha_starter} and \eref{fluc_again} are {\it literally the same}.

\subsection{Adding in the Field Theory}

In Section \ref{main_eg} we outlined three ways of looking at the complex structure stabilization induced by a holomorphic vector bundle. In the proceeding paragraphs, we have seen that for the class of extension bundles defined by \eref{example_bundle} the first two of these approaches are equivalent. It is natural to ask whether anything further can be said about the effective field theory? To this end, we point out here several correspondences between the effective field theory description and the methods described above.

Recall that the F-terms corresponding to $F_{ab}$ in the effective
field theory are governed by the superpotential given to lowest order
in \eref{mrfterm} by \beq\label{sup_again} W =
\lambda_{ij}(\mathfrak{z}_0)C_{+}^{i}C_{-}^{j} \ , \eeq where the
singlet fields are defined by \beq\label{singlet_c} C_{+}^{i} \in
H^1(X, \cL^2)~~~~,~~~C_{-}^{j} \in H^1(X,{\cL^{\vee}}^2) \eeq and
$H^1(X,{\cL^{\vee}}^2)\simeq H^2(X, \cL^2)$ by Serre Duality
\cite{AG}. The F-term that can potentially break supersymmetry is
given by \beq\label{f_ness} \frac{\partial W}{\partial
  C_{-}^{j}}=\lambda_{1j}(\mathfrak{z}) \langle C_{+}^{1} \rangle \ .
\eeq It is only when the complex structure moduli are constrained to
the locus for which $\lambda_{1j}(\mathfrak{z}_{0})=0$ that this
F-term can vanish and we have a vacuum.

To solve the effective field theory of Section \ref{main_fieldtheory}
directly, it is essential to determine the function
$\lambda_{1j}(\mathfrak{z})$ in \eref{sup_again} and for which values
$\mathfrak{z}_0$ it vanishes. By definition, on this locus the
$C$-fields in \eref{singlet_c} are massless modes of the
four-dimensional theory. When $\lambda_{1j}(\mathfrak{z})$ vanishes,
the fields in \eref{singlet_c} correspond to harmonic elements of the
given {\it non-trivial} cohomology groups.  However, by definition,
finding the locus $\mathfrak{z}_0$ for which the $C$-fields correspond
to elements of their respective non-trivial cohomologies is precisely
the content of the first-order fluctuation analysis given in
\eref{fluc_again} above. That is, by definition, where the cohomology
exists the $C_{+,-}$ are massless fields and where it does not, they
are massive. As we have seen, this is equivalent to the Atiyah
computation. We conclude from the results of this section that the
effective field theory analysis is identical to Atiyah and ``jumping''
formalisms.

There is further structure that can be observed from the F-term in
\eref{f_ness}. Note that the number of constraints on the complex
structure moduli arising from the vanishing of $F_{C_{-}}$ in
\eref{f_ness} can be at most $h^1(X, {\cL^{\vee}}^2)$, the number of
$C_{-}$ fields\footnote{Note that the number of moduli stabilized need
  not exactly be the number of $C_{-}$ fields, since even if $\alpha$
  in \eref{alpha_starter} is non-trivial, it need not be
  surjective.}. However, by inspection, this space is Serre dual to
$H^2(X, \cL^2)$, the image of the Atiyah map in
\eref{alpha_starter}. Since $\textnormal{dim}(\textnormal{Im}(\alpha))
\leq h^2(X, \cL^2)$, it is clear that the field theory and the Atiyah
computation give us the same upper bound on the number of moduli
stabilized for a given bundle $V$ in \eref{example_bundle}. In
addition, the form of the superpotential in \eref{sup_again} makes it
clear that the mass term can only remove even numbers of $C_{+}$ and
$C_{-}$ fields, in agreement with the jumping calculation of
\eref{koszul_with_feeling} and \eref{fluc_again} which tells us that
the cohomology of $\cL^2$ can only jump in a way that preserves the
index, $\textnormal{Ind}(\cL^2)$.

Importantly, although all three approaches carry the same content, they are not all equally difficult to analyze. Without a knowledge of the ``jumping locus", it would be hard to directly calculate the coefficient $\lambda(\mathfrak{z}_0)$ in \eref{sup_again} via any standard tools in the effective field theory. And while the Atiyah computation can be shown to carry the same content as the jumping locus, it too is seemly much more complicated at first glance. In particular, determining which class $\alpha \in H^1(X, \cL^{2}\otimes V \otimes V^{\vee})$ corresponds to the physical field strength, $F^{1,1}$ (and, hence, the initial vacuum) is a notoriously difficult problem. Thus, we present the results of this section as an important example of ways in which the ``bottom up" analysis of bundle holomorphy can provide an efficient method to study complex structure stabilization.

\section{An Explicit $SU(2)$ Example}\label{our_eg}

In this section, we give an explicit example of the techniques developed in this paper by presenting a holomorphic bundle which stabilizes most of the complex structure moduli of its base. Specifically, in this example, the constraints arising from bundle holomorphy stabilize all but two of the complex structure moduli -- fixing $80$ out of $82$ degrees of freedom. In addition, since one must include Wilson lines in realistic heterotic compactifications, we also consider the same calculation combined with a freely acting discrete automorphism, $\Gamma$. In this case, the quotient bundle on $X/\Gamma$ stabilizes $10$ out of $11$ complex structure moduli (the technical details of this last calculation can be found in Appendix \ref{comp_detailsec}).

As discussed previously, there are three equivalent ways of
determining the number of complex structure moduli stabilized by a
holomorphic vector bundle satisfying the Hermitian-Yang-Mills
equations: 1) by computing the image of the Atiyah class as in
\eref{atiyah_coh}, 2) by using something inherently complex structure
dependent in a given type of bundle construction to compute where in
moduli space such a bundle can be holomorphic (a ``jumping''
calculation) and 3) by solving F-terms in the four-dimensional
effective field theory.  In this section, we investigate an explicit
geometry following the first approach. We will directly infer for
which values of the complex structure the bundle can be kept
holomorphic by constructing a vector bundle which manifestly depends
on the complex structure of its base.

To begin, consider the complete intersection Calabi-Yau threefold defined by
\begin{eqnarray}
\label{eg1cy3}
X=\left[ \ba{c |c }
\mathbb{P}^2 & 3 \\
\mathbb{P}^2 & 3
\ea \right] \ ,
\end{eqnarray}
with $h^{2,1}=82$ complex structure moduli. For this manifold, the number of K\"ahler moduli is given by $h^{1,1}=2$. If we expand the K\"ahler form $\omega$ in a basis of harmonic $\{1,1\}$-forms as $\omega=t^{r}\omega_{r}$ with $r=1,2$, the K\"ahler moduli $t^r$ take values such that $t^r >0~\forall~r$. The tangent bundle for this space is defined by the pair of sequences
\bea\label{tangentbundle}
&&0 \to {\cO_{X}}^{\oplus 2} \to \cO(1,0)^{\oplus 3} \oplus \cO(0,1)^{\oplus 3} \to T{\cal A} \to 0 \ , \\
&&0 \to TX \to T{\cal A} \to \cO(3,3) \to 0
\eea
where both sequences are restricted to $X$ and $T{\cal A}$ is the tangent bundle to the ambient space $\mathbb{P}^2 \times \mathbb{P}^2$. Using \eref{tangentbundle}, it is straightforward to show that the complex structure moduli of $X$ are given by
\bea
H^{2,1}(X)=H^1(X,TX)=H^0(X,\cO(3,3))/H^0(X,T{\cal A}|_{X})\ , 
\label{b1}
\eea
where
\bea
H^0(X,T{\cal A}|_{X})= (H^0(X,\cO(1,0)^{\oplus 3}\oplus \cO(0,1)^{\oplus 3}))/H^0(X,{\cO_{X}}^{\oplus 2}) \ .
\label{b2}
\eea
Stated simply, the complex structure moduli of $X$ are given by the number of degree $\{3,3\}$ polynomials that define the hypersurface in \eref{eg1cy3}, modulo a series of $GL(3, \mathbb{C})$ transformations (corresponding to coordinate redefinitions in the projective spaces, $\mathbb{P}^2$ and $\mathbb{P}^2$, in \eref{eg1cy3}). As a result, in the following, we will frequently make a choice of complex structure moduli by specifying $p_0$, a defining bi-degree $\{3,3\}$ polynomial in $\mathbb{P}^2 \times \mathbb{P}^2$ (where the appropriate $GL(3,\mathbb{C})$ redundancies have been taken into account).

Over this manifold, we define the following indecomposable $SU(2)$ bundle by extension,
\beq\label{Vdef}
0 \to\cO(-3,3) \to V \to \cO(3,-3) \to 0 \ .
\eeq
This bundle satisfies the anomaly cancellation condition \eref{anomaly1}. Further, we choose (and hold constant throughout this section) a point in K\"ahler moduli space for which the indecomposable $SU(2)$ bundle is slope stable. For the K\"ahler moduli $t^r$ defined above, this is the subspace of the positive quadrant such that $\mu(\cO(-3,3))=c_1(\cO(-3,3))^{r}d_{rst}t^{s}t^{r} <0$. Here, the triple intersection numbers are given by $d_{112}=d_{122}=3$. Note that for this choice of the K\"ahler moduli, the {\it only supersymmetric configuration} for the bundle defined by \eref{Vdef} is as an indecomposable rank 2 bundle. The split extension ${\cal L} \oplus {\cal L}^{\vee}$ {\it is not poly-stable} in this region of K\"ahler moduli\footnote{Note that, by definition, a poly-stable bundle is a direct sum of properly stable bundles, $V=\bigoplus_i V_i$, all with the same slope $\mu(V)=\mu(V_i)~\forall i$ \cite{Huybrechts}.} space since $\mu({\cal L}) \neq \mu({\cal L}^{\vee})$ and, hence, its connection does not solve $g^{a\bar{b}}F_{a\bar{b}}=0$. Moreover, deep within the stable chamber of the K\"ahler cone, the supersymmetric connection on the indecomposable $SU(2)$ bundle \eref{Vdef} is {\it not infinitesimally close} to the split $S[U(1) \times U(1)]$ valued connection of ${\cal L}\oplus {\cal L}^{\vee}$. Rather, it's ``off diagonal" $SU(2)$-valued components are finitely large and cannot be set to zero by a small change, $\delta A$.

The extension class of $V$ is an element of $Ext^1({\cal L}^{\vee}, {\cal L})=H^1(X,{\cal L}^2)$. However, direct computation yields 
\beq\label{cohom}
H^1(X,{\cal L}^2)=H^1(X, \cO(-6,6))=0~~~\text{for generic}~p_0 \ .
\eeq
Since the space of extensions is generically trivial, it is clear that the indecomposable $SU(2)$ bundle $V$ in \eref{Vdef} cannot be defined for generic values of the complex structure moduli. However, as we saw in previous sections, cohomology can ``jump" over certain higher-codimensional loci in complex structure moduli space. Just such a jumping is possible here and we will be interested in determining that locus.

As discussed in Subsection \ref{jumpingsection}, the cohomology of a
line bundle, ${\cal L}^2$, on the threefold in \eref{eg1cy3} is
determined in terms of line bundles on the ambient multi-projective
space ${\cal A}= \mathbb{P}^2\times\mathbb{P}^2$ via the Koszul
complex \cite{AG} \beq\label{koszully2} 0 \to \cN_{\cal A}^{\vee}
\otimes {\cal L}_{\cal A}^2 \stackrel{p_0}{\to} {\cal L}_{\cal A}^2
\to {\cal L}^2 \to 0 \ , \eeq where $\cN_{{\cal A}}=\cO(3,3)$ defines
the normal bundle to $X$.  To determine the cohomology $H^{*}(X, {\cal
  L}^2)$, we take the long exact sequence in cohomology associated
with \eref{koszully2}. We find that the cohomology $H^1(X, \cO(-6,6))$
in \eref{cohom} is defined in terms of the cohomology of line bundles
over the ambient space ${\cal A}$ via the short exact sequence
\beq\label{koszul_cohom} 0 \to H^1(X, \cO(-6,6)) \to
H^2(\cA,\cO(-9,9)) \stackrel{p_0}{\to}H^2(\cA, \cO(-6,6)) \to H^2(X,
\cO(-6,6)) \to 0 \ .  \eeq As a result, for special values of the map
$p_0$ in the short exact sequence above, 
\beq\label{ker_eq}
\textnormal{Ker}(p_0)=H^1(X,\cO(-6,6)) \neq 0~.  
\eeq 
Using the
Bott-Borel-Weil polynomial representations for the ambient space
cohomology groups \cite{Anderson:2008ex} (see Appendix \ref{comp_detailsec}), it is possible to explicitly
formulate and solve the equation for $\textnormal{Ker}(p_0)$ in terms
of the defining $\{3,3\}$ polynomials. Direct computation
\cite{us_to_appear,Gray:2008zs} demonstrates that on a two-dimensional
locus in complex structure moduli space, the space of extensions
``jumps" to $Ext^1({\cal L}^2)=H^1(X,\cO(-6,6))=180$.

Now choose a background somewhere in this two-dimensional sub-locus of
complex structure moduli space for which the non-trivial extension,
\eref{Vdef}, can be defined. Given a well-defined starting point
satisfying the Hermitian Yang-Mills equations, one can now ask: what
happens as we vary the complex structure moduli away from our starting
point within the two-dimensional locus?  As we vary the complex
structure so that $\textnormal{Ker}(p_0) \to 0$, the defining
cohomology class goes to zero. More explicitly, as described in
Section \ref{jumpingsection}, we find that the form describing the
extension class $\beta \in H^1(X,\cO(-6,6))$ defining $V$ does not
vanish as we leave the two-dimensional locus. Instead, under a change
of complex structure it is no longer a harmonic $(0,1)$-form (that is,
an element of $H^1(X,V\otimes V^{\vee})$) with respect to the new
definition of the holomorphic coordinates! In other words, $V$ is no
longer a holomorphic bundle. As a result, by determining the locus in
complex structure moduli space where the cohomology defining the
extension class is trivial/non-trivial, we are actually determining
for which values of the complex structure moduli it is possible to
satisfy $F_{ab}=0$.  We can perform the fluctuation analysis described
in Section \ref{jumpingsection}, Eq. \eref{jump_fluc} to decide for
which values $\delta p$ can $\textnormal{Ker}(p_0 + \delta p) \neq
0$. That is, we solve \beq\label{fluct_again} (p_0 + \delta p)(k_0
+\delta k)=0 \eeq for any $\delta k, \delta p$. By direct calculation,
we find that the bundle $V$ in \eref{Vdef} stabilizes $80$ out of the
$82$ complex structure moduli on $X$.

As was demonstrated in Section \ref{equivalent_approaches}, for this
simple class of bundles the calculation above is literally the same
polynomial multiplication problem (although in different form) that
one must solve to do the computation from the point of view of the
Atiyah sequence. However, as a non-trivial consistency check of the
results of \eref{fluct_again}, it is worth recalling that from the
Atiyah sequence \beq \ldots \to H^1(X,{\cal Q}) \stackrel{d\pi}{\to}
H^1(X,TX) \stackrel{\alpha}{\to} H^2(X,V\otimes V^{\vee}) \to \ldots \
, \eeq the number of stabilized complex structure moduli is given by
the dimension of $\textnormal{Im}(\alpha)$. In this example, at the
holomorphic starting point (where $\textnormal{Ker}(p_0) \neq 0$),
\beq h^2(X,V\otimes V^{\vee})=h^1(X,V\otimes
V^{\vee})=h^1(X,\cO(-6,6))+h^1(X,\cO(6,-6))-1=359, \eeq using Serre
duality. Thus, it is clear that stabilizing $80$ of the complex
structure moduli is consistent with the existence of an appropriate
image $\textnormal{Im}(\alpha) \subset H^2(X, V\otimes V^{\vee})$ for
the Atiyah map, as expected.

\subsection{Equivariant Structures and $X/(\mathbb{Z}_3\times \mathbb{Z}_3)$}\label{equiv_downstairs}
In this subsection, we point out that there is one more tool at our disposal. For realistic phenomenology, one wants to include Wilson lines and, hence, we need a manifold with $\pi_1(X) \neq 0$. As a result, one could perform the complete analysis of the previous subsection ``downstairs" on the quotient manifold, $X/\Gamma$, where $\Gamma$ is a freely acting discrete automorphism of $X$. In general, such quotient manifolds have fewer complex structure moduli than the ``upstairs" covering spaces $X$ and, as a result, we would hope to stabilize even more moduli.

The Calabi-Yau threefold in \eref{eg1cy3} admits a freely acting $\mathbb{Z}_3\times \mathbb{Z}_3$ symmetry. If we denote the coordinates on $\mathbb{P}^2\times \mathbb{P}^{2}$ by $\{x_i,y_i\}$, where $i=0,1,2$, then a freely acting $\mathbb{Z}_3 \times \mathbb{Z}_3$ symmetry is generated by \cite{Candelas:2007ac}
\begin{equation}\label{z3z3}
\begin{array}{llll}
{\mathbb{Z}_3}^{(1)}&:&x_k \to x_{k+1},&y_k \to y_{k+1}\\
{\mathbb{Z}_3}^{(2)}&:&x_k \to \alpha^k x_{k},&y_k \to \alpha^{-k} y_{k}
\end{array}
\end{equation}
with $\alpha=\exp (2\pi i/3)$. 
As shown in Ref.~\cite{Candelas:2007ac}, the most general bi-degree $(3,3)$ polynomial invariant under the above symmetry is given by
\begin{eqnarray}\label{bicubic_poly}
p_{(3,3)}&=&A_{1}^{k,\pm} \sum_{j} x_{j}^{2}x_{j\pm 1}y^{2}_{j+k}y_{j+k\pm1}+A_{2}^{k}\sum_{j}x^{3}_{j}y^{3}_{j+k}+A_{3}x_1x_2x_3\sum_{j}y^{3}_j\nonumber\\
&&+A_{4}y_1y_2y_3\sum_{j}x^{3}_{j}+A_5x_1x_2x_3y_1y_2y_3 \ ,
\end{eqnarray}
where $j,k=0,1,2$ and there are a total of $12$ free coefficients, denoted by $A$ with various indices. One combination of coefficients drops out due to the freedom of an overall scaling, leaving $h^{2,1}(X/(\mathbb{Z}_3\times \mathbb{Z}_3))=11$.

Vector bundles on $X$ descend to vector bundles on $X/\Gamma$ if and only if they admit an equivariant structure \cite{equivariance,Donagi:2003tb}. An equivariant structure is a consistent lifting of the isometry $\Gamma$ to the bundle which commutes with the projection $\pi: V \to X$. For the geometry at hand, the line bundles $\cO(-3.3)$ are equivariant with respect to the $\mathbb{Z}_3\times \mathbb{Z}_3$ action. By construction, the $SU(2)$ bundle $V$ in \eref{Vdef} has an induced equivariant structure \cite{Anderson:2009mh} from $\cO(-3,3)$ and its dual, and descends to a rank two bundle, ${\hat V}$, on the quotient space. The cohomology of $\hat V$ on $X/\Gamma$ is simply the subset of $H^*(X, V)$ which is invariant under the group action. 

Repeating the fluctuation analysis of \eref{fluct_again} ``downstairs"
in invariant cohomology, leads to one less free modulus. That
is, the quotient bundle $\hat V$ on $X/(\mathbb{Z}_3 \times
\mathbb{Z}_3)$ stabilizes $10$ out of the $11$ degrees of freedom
remaining in \eref{bicubic_poly} (see Appendix \ref{comp_detailsec}). Thus, there is only one complex
structure modulus remaining in the presence of the holomorphic bundle
$\hat V$. A direct computation \cite{us_to_appear} shows that the
Calabi-Yau threefold defined at this one-dimensional locus in complex
structure moduli space remains smooth.

\subsection{4D Field Theory for the Explicit Example}

Finally, looking briefly at the third approach, one can gain some physical intuition for the chosen geometry. The effective field theory is the one presented in Section \ref{main_fieldtheory}. For the bundle $V$ in \eref{Vdef}, the superpotential of the $E_7$ GUT theory is given to lowest order by
\beq
W \sim \lambda({\mathfrak z})_{ij}C_{+}^iC_{-}^j+ \ldots
\eeq
where $C^{i}_{+},C^{j}_{-}$ are the four-dimensional $E_7$-singlet fields associated with $H^1(X, \cO(-6,6))$ and $H^1(X, \cO(6,-6))$ respectively. As discussed in Section \ref{main_fieldtheory}, the quadratic mass term in the potential must vanish on the locus in complex structure moduli space where the space of extensions $H^1(X, \cO(-6,6))$ is non-trivial (that is, $V$ is a holomorphic vector bundle), since here $H^1(V\otimes V^{\vee})\neq 0$ parametrizes the {\it massless states} of the effective theory. That is, for ${\mathfrak z}_0$ in the two-dimensional locus
\beq
\lambda({\mathfrak z}_0)=0~.
\eeq
Further, as shown in Section \ref{main_fieldtheory}, since $V$ in \eref{Vdef} is defined by a non-trivial vev $C_{+} \in H^1(X, \cO(-6,6))$, the condition that $F_{ab}=0$ corresponds to the vanishing of the F-term,
\beq\label{ftermy}
\frac{\partial W}{\partial C_{-}^{j}}= \lambda_{j1}({\mathfrak z})\langle C_{+}^{1} \rangle \ .
\eeq
Note that the number of constraints arising from \eref{ftermy} is counted by the number of $C_{-}$-fields. As demonstrated in \eref{restricted_alpha} and \eref{restricted_im}, for the bundles $V$ in this section the image of the Atiyah map is $H^2(X, \cL^{2})$ which, by Serre duality, is again the space of $C_{-}$-fields! Thus, the field theory and the geometry encode the same information --  as we expect.

Finally, note that for the bundle $V$ in \eref{Vdef} the $C_{-}$
fields are massless in the holomorphic vacuum. Hence, it is clear
that, when we consider the F-term associated with $\frac{\partial
  W}{\partial \mathfrak{z}^a}$, this geometry is an example of the
higher-order lifting described at the end of Section
\ref{main_fieldtheory}.

\section{Other Examples}\label{other_section}

In this section, we briefly explore other examples of holomorphic vector bundles which can stabilize the complex structure. Specifically, we illustrate how it is possible that something more complex than a single line bundle cohomology can control the holomorphy of a vector bundle.

Having shown in Section \ref{equivalent_approaches} that the field theory, ``jumping Ext" and the Atiyah approaches are equivalent for the class of examples given in \eref{example_bundle}, we henceforth use the formalism that is most convenient for analyzing the complex structure of a given vacuum. One of the goals of this section is to demonstrate that examples of complex structure stabilizing vector bundles are readily available and that their dependence on complex structure is easily analyzed.

\subsection{``Jumping" Tri-linear Couplings}
Here, we consider another example of a heterotic compactification in which the vanishing of F-terms associated to $F_{ab}$ leads to the stabilization of the complex structure. This example will illustrate that the key geometric ingredient needed to define a holomorphic bundle can be more complicated than the simple line bundle cohomology \eref{the_best_ext} controlling the holomorphy of the bundle in \eref{example_bundle}.

Consider the $SU(3)$ bundle $V$ defined by the two extension sequences
\bea\label{double_ext}
0 \to L_1 \to U \to L_3 \to 0 \ ,\\
0 \to U \to V \to L_2 \to 0\ . \nonumber
\eea
This bundle is slope-stable as long as the K\"ahler moduli and the line bundles, $L_{1}, L_{2}$ and $L_{3}$, are chosen so that $\mu(L_1), \mu(L_2) <0$. In addition, the second Chern classes of the bundle must be such that anomaly cancellation, Eq. \eref{anomaly1}, is satisfied (see Section \ref{hidden_section} for further discussion).

Of the three approaches to complex structure stabilization illustrated in Section \ref{main_eg}, here we use  the effective field theory \cite{Anderson:2010tc,Anderson:2010ty} and geometric methods associated with the bundle \eref{double_ext}. As in Section \ref{main_fieldtheory}, we analyze the effective field theory near the region in K\"ahler/bundle moduli space for which the bundle decomposes as $V \to L_1 \oplus L_2 \oplus L_3$. At this locus in moduli space, the structure group of $V$ changes from $SU(3)$ to $S[U(1)\times U(1) \times U(1)] \simeq U(1)\times U(1)$. These anomalous $U(1)$ symmetries are self-commuting within $E_8$ and give rise to a visible $E_6 \times U(1) \times U(1)$ gauge symmetry in the four-dimensional theory. The fields charged under $E_6$ will all be set to zero in the vacuum and, as in previous sections, only the singlet fields will be of interest to us. As before, the $E_6$ singlets arising from $H^1(X, V \otimes V^{\vee})$ acquire $U(1)$ charges under these enhanced symmetries. 

We assume for this discussion that the following charged bundle moduli are present
\begin{align}{\bf 1}_{1/2,3} &= {C_1}^{i} \in H^1(X, L_2 \otimes {L_1}^{\vee}) \ ,\nonumber \\
{\bf 1}_{1/2,-3} &= {C_2}^{a} \in H^1(X, L_1 \otimes {L_3}^{\vee})\ , \label{u1charges}
 \\
{\bf 1}_{-1,0} &= {C_3}^{\beta} \in H^1(X, L_3 \otimes {L_2}^{\vee}) \ .\nonumber
\end{align}
The two anomalous $U(1)$ symmetries give rise to two D-terms \cite{Anderson:2010tc} of the same form as \eref{mrdterm},
\begin{align}
&D^{U(1)_{1}} \sim \frac{\mu(L_1)}{{\cal V}} -{Q^{1}}_{i{\bar j}} {C_1}^{i} {{\bar C}_{1}}^{{\bar j}}-{Q^{1}}_{a{\bar b} }{C_2}^{a}{ {\bar C}_2}^{{\bar b}}-{Q^{1}}_{\alpha {\bar \beta}} {C_3}^{\alpha} {{\bar C}_3}^{{\bar \beta}} \ , \\
&D^{U(1)_{2}} \sim \frac{\mu(L_2)}{{\cal V}} -{Q^{2}}_{i{\bar j}} {C_1}^{i} {{\bar C}_{1}}^{{\bar j}}-{Q^{2}}_{a{\bar b} }{C_2}^{a}{ {\bar C}_2}^{{\bar b}}-{Q^{2}}_{\alpha {\bar \beta}} {C_3}^{\alpha} {{\bar C}_3}^{{\bar \beta}} \ ,
\end{align}
where $Q^{1,2}$ denote the $U(1)$ charges of the fields with respect to each of the enhanced $U(1)$-symmetries. These are given in the subscripts on the left-side of equations \eref{u1charges}. We will set these D-terms to zero in the vacuum by giving non-vanishing vevs to one each of the fields $C_2$ and $C_3$. That is,
\begin{align}
&D^{U(1)_{1}} \sim \frac{\mu(L_1)}{{\cal V}} +|\langle C_3 \rangle|^2 -\frac{1}{2}|\langle C_2 \rangle|^2=0 \ ,\\
&D^{U(1)_{2}} \sim \frac{\mu(L_2)}{{\cal V}} +3|\langle C_2 \rangle|^2=0 \ .\label{2dterm2}
\end{align}
By this vev choice, we define the two non-trivial extensions in \eref{double_ext}. The fact that the D-terms can be satisfied is in agreement with the fact that the $SU(3)$ bundle in \eref{double_ext} is slope-stability as long as $\mu(L_1) <0$ and $\mu(L_2) <0$. 

To complete the effective theory, one must consider the holomorphic
superpotential.  This is given, to leading order, by
\beq\label{superpot} W =
\lambda_{ia\alpha}(\mathfrak{z}){C_1}^{i}{C_2}^{a}{C_3}^{\alpha} \ .
\eeq In the vacuum, $\langle C_2 \rangle \neq 0$, $\langle C_3 \rangle
\neq 0$ and, hence, to set the F-terms \beq \frac{\partial W}{\partial
  C_2} = \lambda(\mathfrak{z}) \langle C_3 \rangle
C_1~~,~~\frac{\partial W}{\partial C_3} \sim \lambda(\mathfrak{z})
\langle C_2 \rangle C_1 \ \eeq to zero, one must choose the remaining
vevs $\langle C_1 \rangle =0$.  As in Section \ref{main_fieldtheory},
for a suitable basis on field space, the potentially non-vanishing
F-term of interest takes the form \beq\label{cool_fterm}
\frac{\partial W}{\partial {C_1}^{i}} =
\lambda_{i11}(\mathfrak{z})\langle {C_2}^{1} \rangle \langle {C_3}^{1}
\rangle \ .  \eeq To preserve $N=1$ supersymmetry in the vacuum, one
must set the F-terms (counted by the $h^1(X,L_2\otimes {L_1}^{\vee})$
$C_1$ fields) to zero, \beq\label{lambdazero}
\lambda_{i11}(\mathfrak{z})=0 \ .  \eeq This constrains us to a
sub-locus in complex structure moduli space, as in Section
\ref{main_fieldtheory}. As argued in Appendix \ref{hom_scaling}, the
homogeneous linear scaling of the superpotential with complex structure
guarantees that \eref{lambdazero} provides a non-trivial constraint on
the complex structure.

From the point of view of the four-dimensional field theory, the complex structure dependence of the Yukawa couplings in \eref{superpot} gives rise to vacuum stabilization of the complex structure via the F-term in \eref{cool_fterm}. It is of interest to ask next how these constraints can manifest themselves in the geometry associated with \eref{double_ext}?

The first extension sequence is governed, once again, by a simple line bundle cohomology, $Ext^1(L_3, L_1)=H^1(X, {L_3}^{\vee}\otimes L_1)$. This non-trivial extension is the geometric realization of the vev choice $\langle C_2 \rangle \neq 0$ in \eref{2dterm2}. But what governs the splitting of the second sequence? The second sequence in \eref{double_ext} is controlled by $Ext^1(L_2, U)=H^1(X, {L_{2}}^{\vee} \otimes U)$ which is, in turn, defined via the exact sequence
\beq
0 \to L_1 \otimes L_{2}^{\vee} \to U \otimes L_{2}^{\vee} \to L_3 \otimes L_{2}^{\vee} \to 0 \ .
\eeq
The long exact sequence in cohomology associated with this is
\begin{align}\label{second_extcoh}
&0 \to H^1(X,L_1 \otimes L_{2}^{\vee} ) \to H^1(X,U \otimes L_{2}^{\vee}) \to  H^1(X,L_3 \otimes L_{2}^{\vee} ) \\ \nonumber
&\stackrel{\rho}{\hookrightarrow} H^2(X,L_1 \otimes L_{2}^{\vee} ) \to \ldots
\end{align}
where the co-boundary map, $\rho$, is \beq\label{rho_def} \rho \in
H^1(X, L_1 \otimes {L_3}^{\vee}) \ .  \eeq It is clear from
\eref{second_extcoh} that one can decompose the space of extensions
$H^1(X,U \otimes L_{2}^{\vee})$ as \beq\label{split_extclass} H^1(X,U
\otimes L_{2}^{\vee}) \simeq H^1(X, L_1 \otimes L_2^{\vee}) \oplus
(\textnormal{Ker}(\rho) \subset H^1(X,L_3 \otimes {L_2}^{\vee})) \ .
\eeq For this geometry to match the field theory vacuum presented
above, it must be that the non-trivial extension arises from the
second term in \eref{split_extclass}; that is, from \beq \langle C_3
\rangle \in H^1(X,L_3 \otimes {L_2}^{\vee}) \ .  \eeq However, it is
clear from \eref{second_extcoh} that such a choice of non-trivial
extension class may not exist. There is an important condition that
must be satisfied; namely, that $\textnormal{Ker}(\rho)$ be non-trivial. Let us
look at this constraint in more detail by considering that, for a
fixed $\rho$ in \eref{rho_def}, \beq H^1(X,L_3 \otimes L_{2}^{\vee} )
\times H^1(X, L_1 \otimes {L_3}^{\vee}) \to H^2(X,L_1 \otimes
L_{2}^{\vee} ) \eeq must have a trivial image for some element of
$H^1(X,L_3 \otimes L_{2}^{\vee} )$. Using Serre duality, one can
multiply both sides by $H^1(X, L_2 \otimes {L_1}^{\vee})$ to obtain
the tri-linear product \beq\label{yoneda} H^1(X,L_3 \otimes
L_{2}^{\vee} ) \times H^1(X, L_1 \otimes {L_3}^{\vee}) \times H^1(X,
L_2 \otimes {L_1}^{\vee}) \to H^3(X, \cO_X) \simeq \mathbb{C} \ .
\eeq In order for any element of $H^1(X, L_3 \otimes L_{2}^{\vee})$ to
be in $\textnormal{Ker}(\rho)$, the product in \eref{yoneda} must map to zero. But,
by inspection, this Yoneda (triple) product as a map into $\mathbb{C}$
is simply the coefficient of \beq \langle C^{1}_3 \rangle\langle
C^{1}_2 \rangle C_1^{i}~.  \eeq That is, it is exactly the Yukawa
coupling, $\lambda_{i11}(\mathfrak{z})$, in \eref{lambdazero} and
\eref{cool_fterm}! The geometric statement of \eref{yoneda} is that we
cannot choose a non-trivial extension $\langle C^{1}_3 \rangle \in
H^1(X,L_3 \otimes {L_2}^{\vee})$ unless its product with $\langle
C_{2}^{1} \rangle$ (the extension class of \eref{rho_def}) and all
$C_1$ fields vanishes. However, by inspection this is simply the
statement that no supersymmetric heterotic vacuum exists unless the
F-term in \eref{cool_fterm} is set to zero -- that is, that the Yukawa
coupling $\lambda(\mathfrak{z})$ satisfies \eref{lambdazero}.

From these two separate analyses of the heterotic vacuum defined by the bundle in \eref{double_ext}, we see that the field theory and the geometry encode the same constraints on complex structure. This stabilization could be explored further by performing an Atiyah computation as in Section \ref{full_atiyah} for the bundle in \eref{double_ext}. However, even without performing this analysis, it is clear that the presence of such an $SU(3)$ bundle constrains the moduli. By arguments similar to those in Section \ref{main_fieldtheory} and equation \eref{box4}, a vacuum fluctuation of \eref{cool_fterm} demonstrates that $\delta \mathfrak{z}$ acquires a mass for directions $\delta \mathfrak{z}_{\perp}$ which move off the locus defined by $\lambda(\mathfrak{z})=0$.

Geometrically, the triple product in \eref{yoneda} depends on complex structure and its image can ``jump" from zero to non-zero values as we vary the complex structure of the base. This class of bundles provides a nice example of ``bottom up" support of the bundle which jumps with complex structure. Unlike the bundle in Section \ref{main_eg}, it is not simply line bundle cohomology which jumps -- rather, in this case, it is the triple product of such cohomology groups. When the moduli are chosen so that Yoneda product vanishes it is possible to define the non-Abelian $SU(3)$ gauge configuration given in \eref{double_ext}. But from both the geometric and field theoretic points of view, when the Yukawa coupling, $\lambda(\mathfrak{z})_{i11}$, given in \eref{lambdazero} and \eref{yoneda}, is non-zero for all $C_3 \in  H^1(X,L_3 \otimes L_{2}^{\vee} )$, then a holomorphic bundle cannot be defined.

\subsection{Monad Bundles and ``Jumping" Homs}
As another example of  a ``bottom up" approaching to studying bundle holomorphy, one can consider different method of constructing holomorphic vector bundles.

Consider the following three term complex, called a ``monad" \cite{monadbook}
\beq\label{gen_monad}
0 \to \bigoplus_k \cO({\bf a}^k) \stackrel{g}{\rightarrow}  \bigoplus_i \cO({\bf b}^k) \stackrel{f}{\rightarrow}  \bigoplus_j \cO({\bf c}^k) \to 0 \ ,
\eeq
where $f \circ g=0$. The complex \eref{gen_monad} defines a holomorphic vector bundle via
\beq\label{monv}
V = \frac{\textnormal{Ker}(f)}{\textnormal{Im}(g)} \ .
\eeq
If we label the sums of line bundles as
\beq
A=\bigoplus_k \cO({\bf a}^k)\ ,~~~  B=\bigoplus_i \cO({\bf b}^k)\ ,~~~ C=\bigoplus_j \cO({\bf c}^k)
\eeq
then the maps $f$ and $g$ take the form
 \beq
 g \in H^0(X, \textnormal{Hom}(A,B))~~, ~~f \in H^0(X,\textnormal{Hom}(B,C)) \ .
 \eeq
 As seen in the previous sections, these cohomologies ``jump" with complex structure. When the complex structure are varied so that components of either of these spaces of global sections go to zero, it is not possible to build the non-Abelian holomorphic vector bundle $V$ in \eref{monv}.
 
 \subsubsection{A Monad Example}
We will consider an explicit example of a ``jumping" monad bundle. Define an indecomposable $SU(2)$ bundle via a monad sequence on the following CY threefold:
\begin{eqnarray}
\label{eg2cy3}
X=\left[ \ba{c |c }
\mathbb{P}^1 & 2 \\
\mathbb{P}^3 & 4
\ea \right] \ .
\end{eqnarray}
The Hodge numbers of this manifold are $h^{1,1}=2$ and $h^{2,1}=86$.

For the monad bundle, we will choose $A$ in \eref{gen_monad} to be zero for simplicity, and consider the two-term monad \cite{monadbook,Anderson:2007nc,Anderson:2008uw, Anderson:2009ge,Anderson:2009mh}
\beq\label{monad_good}
0 \to V \to \cO(2,0)\oplus \cO(-1,2)^{\oplus 2} \stackrel{f}{\rightarrow} \cO(0,4) \to 0 \ .
\eeq
This bundle is slope-stable in the region of K\"ahler moduli space defined by $c_1(\cO(3,-4))^{r}d_{rst}t^{s}t^{t} <0$. The K\"ahler cone of \eref{eg2cy3} is defined by $t^r >0~\forall r$, and the non-trivial triple intersection numbers are given by $d_{122}=4, d_{222}=2$. In addition, $V$ satisfies the anomaly cancellation condition \eref{anomaly1}. The defining polynomial $f$ in \eref{monad_good} is an element of the space $H^0(X, \textnormal{Hom}(B,C))$ and, in particular, its first component, mapping $\cO(2,0) \to \cO(0,4)$, is an element of
\beq\label{mon_hom}
f_1 \in H^0(X, \cO(-2,4)) \ .
\eeq
The line bundle $\cO(-2,4)$ is non-ample and for general values of the complex structure, $h^0(X,\cO(-2,4))=0$. However, for special choices of the defining bi-degree $\{2,4\}$-polynomial, the space of global sections in \eref{mon_hom} can ``jump" so that $h^0(X,\cO(-2,4))=1$. For this locus in complex structure moduli space, the indecomposable $SU(2)$ bundle in \eref{monad_good} can be defined.

To quickly analyze the possibilities of such a bundle for stabilizing the complex structure moduli, we need not undertake the full Atiyah analysis of Section \ref{full_atiyah}. Instead, we can simply analyze where the cohomology in \eref{mon_hom} can jump to a non-zero value. The analysis is very similar to that done for a jumping extension class in Section \ref{main_eg}. We consider here the Koszul sequence
\beq\label{koszul_hom}
0 \to H^0(X, \cO(-2,4)) \to H^1({\cal A}, \cO(-4,0)) \stackrel{p_0}{\rightarrow} H^1({\cal A}, \cO(-2,4)) \to H^1(X, \cO(-2,4)) \to 0 \ .
\eeq
The fluctuation equation in this case takes the form
\beq
(j_0+\delta j)(p_0+\delta p)=0 \ ,
\eeq
where $j_0$ is an initial element of $\textnormal{Ker}(p_0)$ and $\delta p \in H^0(A, {\cal N})$ is in the redundant basis for complex structure moduli space given in \eqref{csdefs}. The fluctuation analysis in this case tells us that $H^0(X, \cO(-2,4))$ is non-vanishing on a $53$-dimensional locus in complex structure moduli space. That is, the holomorphy of the $SU(2)$ bundle in \eref{monad_good} stabilizes $33$ out of $86$ complex structure moduli for the CY threefold in \eref{eg2cy3}. By direct computation (see \cite{Anderson:2008uw}) of the bundle moduli for the monad in \eref{monad_good} we find that $h^1(X,V\otimes V^{\vee})=h^2(X, V\otimes V^{\vee})=36$ which means that the image of the Atiyah map is bounded from above by $36$. This is consistent with the result of $33$ moduli stabilized indicated by the above $\textnormal{Hom}(B,C)$-fluctuation.

The result of the fluctuation analysis above is clearly a lower bound
on the dimension of $\textnormal{Im}(\alpha)$ in
\eref{real_moduli}. Since the holomorphic bundle in \eref{monad_good}
cannot be defined when the map $f_1$ in \eref{mon_hom} is not defined,
it is certain in these cases that a direct Atiyah calculation would
reflect the stabilization imposed by the jumping of \eref{mon_hom}. In
general however, the type of analysis given above is only a lower
bound on the number of moduli stabilized. It is possible that the
Atiyah computation would reveal more subtle constraints on complex
structure. As in Section \ref{equivalent_approaches}, more work would
need to be done in this case to understand the relationship (and
possible equivalence) of the fluctuation analysis given above and the
Atiyah computation in Section \ref{full_atiyah}.



We have presented the examples in this section to demonstrate that is possible to easily construct vector bundles which are holomorphic only for higher co-dimensional loci in complex structure moduli space. As shown in previous sections, it is possible to take several approaches to analyzing the constraints on holomorphy. The most direct and systematic method is a ``top down" approach -- given a vector bundle, $V$, and a holomorphic starting point in the total moduli space, we can use the Atiyah analysis of Section \ref{full_atiyah} to decide how many complex structure moduli of $X$ are constrained by the presence of $V$. This approach is direct and will yield the necessary information. However, it is  frequently a difficult and computationally intensive calculation. Adding to this difficulty is the sensitivity of the answer to the initial starting point for the deformation. Frequently interesting structure exists, but it is hard to know where to start the calculation(see \cite{us_to_appear} for a deeper discussion of this issue and computational tools). 

In this section, we have given several examples of the ``bottom up"
approach described in previous sections.  While this analysis was
proven to be fully equivalent to the Atiyah calculation for the class
of bundles given in Section \ref{main_eg}, in this section we simply
use the bottom-up approach to illustrate, in several different
settings, that it is possible to rapidly gain information about the
holomorphic structure of a class of vector bundles by considering the
way that their central ``support'' (such as the Yoneda product in
\eref{yoneda} and the non-trivial morphisms in \eref{mon_hom}) varies
with the complex structure.

It should be pointed out that the types of examples given in this
section are possible for all known systematic constructions of
holomorphic vector bundles -- including not only extension and monad
bundles defined above but, also, bundles defined by the Spectral Cover
\cite{Friedman:1997ih}, Serre \cite{AG} and the Maruyama
\cite{maruyama,Distler:1987ee} constructions. For any known
construction of vector bundles, one can ask the same questions: {\it
  how does the geometric data supporting the definition of the vector
  bundle vary with complex structure?} By systematically addressing
this question, ``jumping" support (such as the extension classes,
defining $\textnormal{Hom}$s and triple-products discussed above) is
straightforward to identify. As the examples in this section
illustrate, if we envision using point-wise holomorphic bundles as a
mechanism to perturbatively stabilize the complex structure in
heterotic theories, the class of such bundles is apparently both
plentiful and versatile.

\section{Kuranishi Maps, Higher-order Obstructions and the Superpotential}\label{higher_order_sec}

\subsection{Deformations and the Superpotential}

In this subsection, we briefly review the correspondence between higher-order deformations of a geometry and higher-order terms in the effective theory. The essential argument here was first given in \cite{Berglund:1995yu} for the standard embedding and deformations of the tangent bundle, $TX$. Below, we discuss the central ideas in a more general context.

Let us begin with deformation theory. As noted in Section \ref{full_atiyah}, the geometric deformations parametrized by elements of the cohomology groups
\beq\label{defspaces}
H^1(X,TX), ~~H^1(X,\textnormal{End}(V)),~~H^1(X,{\cal Q})
\eeq
that appear as massless degrees of freedom in the four-dimensional effective theory are only {\it first-order} results. That is, these degrees of freedom are only allowable deformations of the geometry to first-order in an infinitesimal expansion. In terms of the effective field theory, these flat directions can, in principle, be lifted by higher-order contributions to the perturbative potential.

We will illustrate the higher-order obstructions to deformations for the bundle moduli, $H^1(X, \textnormal{End}(V))$. However, the results are entirely analogous for the the other two deformation spaces of interest, $H^1(X, TX)$ and $H^1(X, {\cal Q})$. Let $X$ be a compact manifold with {\it fixed complex structure} and $V \stackrel{\pi}{\rightarrow} X$ a holomorphic vector bundle over $X$.  Consider elements $\delta A \in H^1(X, \textnormal{End}(V)))$ corresponding to first-order deformations of the connection on the vector bundle $V$. Such a deformation can be expanded in terms of a vector parameter, $\epsilon$, as
\beq
\delta A(\epsilon)=\epsilon^i a_i + \epsilon^i \epsilon^j a_{ij} + \ldots
\eeq
where $i=1, \ldots h^1(X, \textnormal{End}(V))$. A change $\delta A$ of the connection induces a deformation of the anti-holomorphic covariant operator from ${\bar{D}}_{0}={\bar{\partial}}+A_{0}$ to
\beq\label{dbar}
{\bar D} = {\bar D_0}+ \delta A (\epsilon) 
\eeq
acting on sections of $V$. An allowable deformation is one which preserves the holomorphic nature of the bundle and, hence, keeps $F^{0,2}={\bar D}^2=0$. Imposing this condition, one finds  to first-order in $\epsilon$ that
\beq\label{firstord}
1^{st} \text{-order}:~~~ a_i \in H^1(X, \textnormal{End}(V))
\eeq
and to second-order that
\beq\label{secondord}
2^{nd}\text{-order}:~~~ {\bar D_0}a_{ij}+[a_i,a_j]=0 \ ,
\eeq
where the brackets denote a commutator as elements of $\textnormal{End}(V)$ and a wedge-product as $(0,1)$-forms. As a result, the bracket is actually symmetric in its arguments. In order for the deformation to exist to second-order, $[a_i,a_j]$ must be ${\bar D_0}$-exact.
This pattern continues order by order. At each order in $\epsilon$ one finds a new potential obstruction, which must be trivial as an element of $H^2(X, \textnormal{End}(V))$ for the deformation to be allowed to that order. 

In deformation theory, these obstructions are quantified by a specific map. Specifically, the above obstructions, at each order in $\epsilon$, form a piece of the map
\beq\label{kuranish}
\kappa_{V}: H^1(X,\textnormal{End}(V)) \to H^2(X,\textnormal{End}(V)) \ ,
\eeq
called the {\it Kuranishi map} \cite{kodaira, kuranishi}. More generally for any compact, complex manifold, $Y$, the first-order deformations live in $H^1(X, TY)$ and any obstructions to a holomorphic deformation live in $H^2(X, TY)$. For the case at hand, if all the obstructions described above vanish, then the deformation is called {\it integrable}.  It is clear that if the cohomology group containing all obstructions vanishes identically, that is, $H^2(X, TY)=0$, then the above deformations will be integrable. Unfortunately, for vector bundles on Calabi-Yau threefolds no such simple vanishing occurs. The obstruction spaces $H^2(X, \bullet)$ are non-vanishing for each of the first-order deformation spaces listed above in 
\eref{defspaces}. Hence, more analysis is needed.
Formally, the image of an integrable deformation under the Kuranishi maps vanishes to all orders and the covariant derivative in \eref{dbar} is well-defined. The space of such integrable deformations is
\beq\label{kuranishi_invim}
\kappa_{V}^{-1}(0) \subset H^1(X,\textnormal{End}(V)) \ .
\eeq

What does this geometric structure correspond to in the effective
field theory? To understand this, recall what one means by a zero-mode
of the four-dimensional theory associated with the connection. A
massless state in this case is determined by an endomorphism valued
$(0,1)$-form $a$, which satisfies a bundle-valued Laplace
equation. That is, $a$ is a harmonic representative of an element of
$H^1(X,\textnormal{End}(V))$. By definition, there exist no quadratic
mass terms for $a$ in the potential. Hence, the $1^{st}$ order
criterion ($a \in H^1(X,\textnormal{End}(V))$) for a geometric
deformation implies that the mass terms for the corresponding physical
field vanishes to second-order in the effective potential.  To
second-order in deformation theory, we must compare the result of
\eref{secondord} with the effective theory. The tri-linear
contribution to the superpotential takes the form \beq\label{yukawa3}
y(a_i,a_j,a_k)= \int_X \Omega \wedge \textnormal{Tr}(a_i[a_j,a_k]) \ .
\eeq When \eref{secondord} holds, one can integrate by parts and show
that \eref{yukawa3} vanishes. That is, when \eref{secondord} is
satisfied the second-order geometric deformations are flat to
third-order in the effective theory. It is expected
\cite{Berglund:1995yu} that this correspondence continues to hold to
higher-order -- a deformation is integrable to $n^{th}$ order
geometrically, if and only if the classical contribution to the
spacetime superpotential vanishes through order $(n+1)$.

In each of the three deformation spaces of interest to us,
$\textnormal{Def}(X), \textnormal{Def}(V)$ and
$\textnormal{Def}(X,V)$, we have already discussed the first-order
results in Section \ref{full_atiyah}. One can now ask what happens at
higher-order?  For Calabi-Yau threefolds, this question was
definitively answered for $\textnormal{Def}(X)$ by Tian and Todorov
\cite{tian,todorov}. They showed that all obstructions to the
first-order deformation space, $H^1(X,TX)$, vanish to
higher-order. Phrased in terms of the relevant Kuranishi map,
$\kappa_X: H^1(X,TX) \to H^2(X,TX)$ and the equivalent results to
\eref{firstord} -\eref{kuranish}, \beq \kappa_{X}^{-1}(0)=H^1(X,TX)~.
\eeq From the arguments above, it is clear that in the absence of a
vector bundle to constrain the base manifold, that the complex
structure moduli would remain flat directions of the perturbative
effective theory to all orders.

Now consider the bundle moduli, $H^1(X,\textnormal{End}(V))$, and
inquire what happens to these deformations at higher-order? In the
early literature (see \cite{tyurin,Vafa:1998yp} for example) exploring
the Kontsevich and Strominger-Yau-Zaslow approaches to mirror
symmetry, it was conjectured that, for deformations of slope-zero
stable holomorphic vector bundles, the deformations in
$H^1(X,\textnormal{End}(V))$ were also protected and integrable to all
orders\footnote{Specifically, under certain circumstances slope-zero
  stable vector bundles on Calabi-Yau threefolds $X$ should correspond
  to special Lagrangian cycles (with flat line bundles on them) on the
  mirror $\stackrel{\sim}{X}$. Since the deformations of smoothly
  embedded special Lagrangians are unobstructed \cite{mclean}, it was
  conjectured that the vector bundle moduli would likewise be
  integrable.} -- as in the Tian-Todorov result for
$H^1(X,TX)$. However, subsequent explorations provided explicit
counter examples to this conjecture
\cite{Thomas:1999tq,Aspinwall:2011us}. In particular, it is known that
there can exist non-trivial images $\kappa_V(a) \in H^2(X,
\textnormal{End}(V))$ under the Kuranishi map associated with
$V$. Hence, these flat directions can certainly be lifted to
higher-order in the effective theory.

Finally, consider the deformation space of real interest in a heterotic compactification, $H^1(X,{\cal Q})$. Since this space receives contributions from both of the deformations described above, it is expected that, in general, obstructions can exist and moduli can be lifted at higher-order. In the following subsection, we use the techniques described above to derive some information about the higher-order obstructions to the simultaneous deformation space.

\subsection{A Bound on the Moduli Space}\label{mod_space_bound}
In this section, we show that the dimension of $\textnormal{Def}(X,V)$
to all orders in deformation theory is bounded from below. We are
interested in the image of the Kuranishi map, $\kappa_{(X,V)}:
H^1(X,{\cal Q}) \to H^2(X,{\cal Q})$. To begin, recall the definition
of the bundle $Q$ given in \eref{atiyah_seq}, \beq\label{atiyah_again}
0 \to \textnormal{End}(V) \to {\cal Q} \to TX \to 0 ~, \eeq and it's
associated long exact sequence in cohomology,
\beq\ba{ccccccccc}\label{boundy}
0&\to& H^1(X,\textnormal{End}(V)) &\to& H^1(X,{\cal Q}) &\to& H^1(X,TX) &\to&  \\
&& \kappa_{V}\downarrow &&\kappa_{(X,V)}\downarrow&&\kappa_{X}\downarrow&&\\
&\hookrightarrow& H^2(X,\textnormal{End}(V)) &\to& H^2(X,{\cal Q}) &\to& H^2(X,TX)&\to&\ldots~~. \\
\ea \eeq By following a similar, but unrelated, argument in
\cite{huybrechtspaper}, we will to prove, to all orders in deformation
theory, that \beq \textnormal{dim}(\textnormal{Def}(X,V)) \geq
h^1(X,TX) \ .  \eeq To see this, first note that since each of the
vertical Kuranishi maps above commutes \cite{huybrechtspaper} with the
maps induced from \eref{atiyah_again}, one can relate $\kappa_{(X,V)}$
to $\kappa_{X}$ and $\kappa_V$. We saw above that by the Tian-Todorov
result \cite{tian,todorov}, $\kappa_X$ has vanishing image to all
orders. As a result, $\kappa_{(X,V)}$ has values in the image of
$\rho:H^2(X, \textnormal{End}(V)) \to H^2(X, {\cal Q})$.  By
\eref{boundy} we have
\begin{align}
  &\textnormal{dim}(\textnormal{Def}(X,V))=\kappa^{-1}_{(X,V)}(0)=h^1(X,
  {\cal Q}) -\textnormal{dim}(\textnormal{Im}(\kappa_{(X,V)})\\
  \nonumber &\geq h^1(X,{\cal Q})
  -\textnormal{dim}(\textnormal{Im}(\rho)) \\\nonumber &=h^1(X,{\cal
    Q})-h^2(X, \textnormal{End}(V))+\textnormal{dim}
  (\textnormal{Ker}(\rho)) \ .
\end{align}
By exactness, this gives 
\begin{align}
  &\textnormal{dim}(\textnormal{Def}(X,V)) \geq h^1(X, {\cal Q}) -
  h^2(X,\textnormal{End}(V))
  +\textnormal{dim}(\textnormal{Im}(\alpha)) \\\nonumber &=h^1(X,
  {\cal Q})-h^2(X, \textnormal{End}(V)) +h^1(X, TX)
  -\textnormal{dim}(\textnormal{Ker}(\alpha)) \\\nonumber
  &=h^1(X,\textnormal{End}(V))+\textnormal{dim}(\textnormal{Ker}(\alpha))-h^2(X,
  \textnormal{End}(V))+h^1(X,
  TX)-\textnormal{dim}(\textnormal{Ker}(\alpha)) \\\nonumber
  &=h^1(X,TX)\nonumber
\end{align}
where $\alpha: H^1(X, TX) \to H^2(X, \textnormal{End}(V))$ is the
familiar Atiyah class. The last equality follows from the fact that
$h^1(X,\textnormal{End}(V))=h^2(X,\textnormal{End}(V))$ for $SU(n)$
bundles on a CY threefold by Serre Duality \cite{AG}.

This bound is consistent with the behavior of all bundles thus far
discussed in this paper. For example, line bundles $L$ on a CY
threefold deform with the base and, hence,
$\textnormal{dim}(\textnormal{Def}(X,L))=h^1(X, TX)$ to all
orders. More generally, it was argued in \cite{huybrechtspaper} that
for any rigid bundle with $H^1(X, \textnormal{End}(V))=0$,
$\textnormal{dim}(\textnormal{Def}(X,V))=h^1(X,TX)$. On the opposite
extreme, suppose that we consider a bundle that is only holomorphic
for isolated points in complex structure moduli space. In this case,
as we observed in Section \ref{full_atiyah}, the holomorphy of the
bundle stabilizes all complex structure moduli and $\alpha: H^1(X, TX)
\to H^2(X, \textnormal{End}(V))$ is surjective. That is, $h^1(X,
\textnormal{End}(V)) \geq h^1(TX)$, consistent with the above
bound. Moreover, in this case, the proof given above guarantees that
if all the complex structure moduli are fixed, at least $h^1(X,TX)$ of
the bundle moduli are integrable deformations to all orders. Phrased
in terms of the effective field theory -- for any $N=1$ theory of this
type in which the complex structure moduli have been fixed, there are
at least $h^1(X,TX)$ flat directions in bundle moduli space {\it to
  all orders in the perturbative theory}.

At first glance, this result might indicate that in our approach to complex structure stabilization we have merely shifted the problem from one moduli sector to another, since we must introduce at least $h^1(X,TX)$ bundle moduli to stabilize the complex structure moduli. However, the situation is better than this for a number of reasons. First, at least in some cases, the bundle moduli space for fixed Kahler and complex structure moduli is compact, so that basically any kind of bundle-moduli dependent potential will lead to their stabilization. It is well-known that non-perturbative effects in the superpotential, such as membrane or world sheet instantons and gaugino condensation have pfaffian pre-factors that are higher degree polynomials in the vector bundle moduli \cite{Buchbinder:2002pr,Buchbinder:2002ic,Buchbinder:2002ji,Donagi:1999jp,Curio:2008cm,Curio:2010hd}. Also, the bundles considered here are in the hidden sector so that the coupling of these moduli to observable sector fields is suppressed. As a practical matter, it is important to observe that for simple hidden sector bundles of the form introduced in Section \ref{main_eg}, the bundle moduli space is very simple to describe \cite{Huybrechts} and may be analyzed more easily in the context of moduli stabilization.

Finally, we should make note of these results in relation to the old
(and now disproved) conjectures mentioned in the previous
subsection. Although the claims about the unobstructedness of bundle
moduli space, $H^1(X, \textnormal{End}(V))$, proved to be false, it is
possible that the above bounds shed some light on what subset of
$H^1(X, {\cal Q})$ could be related to un-obstructed (special
lagrangian) geometry under mirror symmetry. It would be interesting to
explore such correspondences further.

\section{A Hidden Sector Mechanism}\label{hidden_section}

From the previous sections, it is clear that the presence of a
holomorphic bundle, $V$, on a Calabi-Yau threefold can highly
constrain the complex structure moduli of the base $X$.  In
particular, by carefully constructing holomorphic bundles (such as the
``jumping" $SU(2)$ extension bundles of Section \ref{main_eg}) it will
be possible to stabilize many or all of the complex structure
moduli. However, it is likely that such constructions will involve
very specific choices of the bundle involved. When one recalls how
difficult it is to construct Heterotic Standard Models (bundles with
the same symmetries and particle spectrum as the MSSM), it is natural
to worry that combining all of these effects into one vector bundle
might be too difficult to accomplish.  Fortunately, within heterotic
M-theory there is more than one gauge bundle to consider. We propose
introducing hidden sector vector bundles (that is vector bundles that
live at the second $E_8$ fixed plane) that will constrain the complex
structure moduli via the mechanism described in this paper. This
allows us to leave the visible sector bundle (in the first $E_8$) free
for model building -- that is, to obtain a realistic gauge group,
spectrum and so on.

To study the possible choices of moduli-stabilizing hidden sector bundles, it is useful to review the conditions that we must place on such a hidden sector. These are
\begin{enumerate}
\item $V$ must be slope poly-stable for some region of the K\"ahler cone.
\item The anomaly cancellation condition \eref{anomaly1} must be satisfied. In full generality, this given by
\beq\label{anomaly_again}
c_2(X)-c_2(V_{visible})-c_2(V_{hidden}) =[W]_{M5}  \ ,
\eeq
linking the second Chern classes of the Calabi-Yau threefold and the visible and hidden sector. 
$[W]_{M5}$ is an effective curve class. Given a visible sector bundle, \eref{anomaly_again} provides a bound on the second Chern class of a hidden sector bundle. 
\item The hidden sector mechanism for complex structure stabilization must be compatible with other phenomenological tools frequently employed in a hidden sector -- such as gaugino condensation. Generically, this has been shown to be possible in \cite{Anderson:2010mh,Anderson:2011cz}.
\end{enumerate}
Given this list of criteria and a fixed visible sector gauge bundle,
it is natural to ask whether the possible choices for complex
structure stabilizing hidden sector bundles can be enumerated? To this
end, we would need to be able to compute the Donaldson-Thomas
invariants \cite{Thomas:1998uj} of the Calabi-Yau manifold $X$ which
count the number of slope-stable, holomorphic vector bundles with a
fixed total Chern class $c(V)=(rank(V),c_1(V),c_2(V),c_3(V))$. It has
been shown that for Calabi-Yau manifolds and fixed $c(V)$, the
associated Donaldson-Thomas invariants are finite
\cite{langer1,langer2}. For our purposes, we know that the rank of $V$
is a maximum of $8$ (and constrained by realistic phenomenology to be
much lower). The first Chern class of $V$ must vanish since its
structure group is a subgroup of $E_8$ and its second Chern class is
bounded from above by \eref{anomaly_again}. Finally, if we hope to add
non-perturbative effects such as gaugino condensation to the hidden
sector, we require the effective theory to be asymptotically free. As
a result, we expect a bound on the third Chern class $c_3(V)$. Since
$c_{3}(V)=\textnormal{Ind}(V)$, this would imply a bound on the number
of hidden sector generations. Thus, for a fixed visible sector gauge
bundle, we would expect that there are a finite number number of
topological choices, $c(V)$, for the hidden sector bundles described
above\footnote{The mathematically inclined reader may wonder if our
  results have any bearing on the Hodge Conjecture (see for example,
  \cite{hodge,lewis}). Specifically, one might ask the following
  question. Let us call a bundle which is not holomorphic for generic
  values of the complex structure of $X$ ``holomorphically
  constrained''. Could it be the case that there exist topological
  classes of vector bundles, parametrized by a total Chern class
  $c(V)$, which are holomorphically constrained for all members of the
  topological class? If so, this would provide a counter-example to
  certain formulations of the Hodge Conjecture. The answer, however,
  is that while such classes of bundles may exist, this is not the
  structure we have seen in the examples presented in this
  paper. Instead, for every complex structure constraining bundle we
  consider there is another member of the topological family which is
  holomorphic for generic values of the complex structure (for
  example, the direct sum $\cL \oplus \cL^{\vee}$ serves this role for
  the bundle defined in \eref{example_bundle}). Thus, our results do
  not contradict any form of the Hodge Conjecture.}. If the relevant
Donaldson-Thomas invariants were to be computed -- see
\cite{Anderson:2010ty} for some techniques to do this -- it should be
possible to describe all relevant hidden sector
bundles. Unfortunately, the computation of high-rank Donaldson-Thomas
invariants is a notoriously difficult problem and one cannot at
present carry out this calculation.

For now, we simply note that on a given Calabi-Yau threefold there may
be many different bundles whose holomorphy will stabilize many or all
of the complex structure moduli. An analysis of an individual
Calabi-Yau threefold may reveal classes of bundles particularly
well-suited to the role of a hidden sector stabilization
mechanism. However, it is important to observe here that even simple
classes of vector bundles, such as the $SU(2)$ extension bundles
discussed in Section \ref{main_eg} and others in Section
\ref{other_section}, can provide good choices of such stabilization
bundles.  In particular, the non-trivial, poly-stable $SU(2)$
extension bundles of Section \ref{main_eg}, \beq\label{hidden_su2} 0
\to {\cal L} \to V \to {\cal L}^{\vee} \to 0 \ , \eeq can be defined
for any Calabi-Yau threefold with $h^{1,1}>1$. Such bundles easily
satisfy the list of constraints above. They are slope-stable in the
region of K\"ahler moduli space for which $\mu(\cL) <0$. Moreover,
their second Chern class can be chosen to be compatible with a wide
range of visible sector bundles. To see this, consider
\beq\label{chern_class} c_2(V)=-ch_2(V)=ch_2({\cal L}) \oplus
ch_2({\cal L}^{\vee})= \frac{c_1({\cal L})^2}{2}+\frac{c_1({\cal
    L}^{\vee})^2}{2}=c_1({\cal L}) \wedge c_1({\cal L}) \eeq where
$r,s=1,\ldots h^{1,1}$. In terms of the triple intersection numbers
$d_{rst}$, we can then write the anomaly cancellation condition as
\beq\label{integer_anomaly}
d_{rst}(c_2(X)-c_2(V_{visible})-c_2(V_{hidden}))^{st} \geq 0~~ \forall
r \ .  \eeq Note that by the stability condition, $H^0(X,{\cal L})=0$,
${\cal L}$ is {\it not an ample line bundle}. As a result, its first
Chern class will generically be a vector with mixed positive/negative
entries. For example, in Section \ref{our_eg}, $c_1({\cal
  L})^r=(-3,3)$. Hence, $d_{rst}c_2(V_{hidden})^{st}$ is generally of
mixed sign and can be taken to be small. Therefore, for a hidden
sector bundle of the type we propose, there will be plenty of room to
fit the visible sector bundle, $V_{visible}$, in the integer bound,
\eref{integer_anomaly}, provided by anomaly cancellation
condition. Finally, as we saw in Section \ref{jumpingsection}, for
appropriate choices of the line bundle ${\cal L}$ the defining $Ext^1$
of the extension sequence can be chosen to ``jump", as discussed in
\eref{jump_fluc}. That is, it is generically possible to find an
extension bundle of this form, whose holomorphic structure will have a
dependence on the complex structure moduli.

Having motivated the utility of this hidden sector mechanism for stabilizing moduli in heterotic compactifications, it is natural to take a step further and ask whether this mechanism can be incorporated into a realistic heterotic moduli-stabilization scenario?
A first exploration of this question was begun in \cite{Anderson:2011cz}. In that paper, the complex structure stabilizing bundles described here were employed in a full heterotic moduli stabilization scenario. It was demonstrated in \cite{Anderson:2010mh,Anderson:2011cz} that bundles of the type \eref{hidden_su2} are not only generically present on Calabi-Yau threefolds, but also completely compatible with other useful hidden sector effects. In particular, we presented a three-stage moduli stabilization scenario for the complex structure, the K\"ahler moduli and the dilaton.

In the first stage, techniques discussed in this work can be employed to stabilize the complex structure moduli by the presence of a vector bundle which
is holomorphic only for an isolated locus in complex structure moduli space. This geometric mechanism can be described by F-term contributions to the effective potential in explicit examples, such as those in Section \ref{main_eg}. For the four-dimensional physics, it is important to note  that the stabilization of the complex structure is achieved without introducing flux. As a result,
the compactification remains a Calabi-Yau threefold and, hence, we are able to retain a considerable mathematical toolkit for analyzing such geometries.

In the second stage, it is possible to use the remaining perturbative condition of slope-stability to restrict the dilaton and K\" ahler moduli. This corresponds to partial D-term stabilization of these fields and is accomplished by adding further ``split" poly-stable bundles to the hidden sector. In \cite{Anderson:2011cz}, we demonstrated that the presence of these D-terms is highly constraining to the effective theory. In particular, the D-terms used in stage 2 are associated with gauging various linear combinations of axions. Any nonperturbative superpotential must be consistent with this.

Finally, in stage 3, we introduce more familiar non-perturbative effects such as gaugino condensation
and membrane instantons. By adding such effects, it was explicitly demonstrated that the non-perturbative mechanisms of stage 3 can complete the stabilization of the geometric moduli (note that the bundle moduli were not considered in \cite{Anderson:2011cz}). A crucial aspect of the scenario described above is that, at the end of stages 1 and 2, the resulting moduli space of vacua is supersymmetric and Minkowski. That is, the unstabilized fields have no potential and the
classical cosmological constant is zero. As a result, this scenario does not suffer from a need to Òfine-tuneÓ the perturbative potential to be small, as arises in some ÒKKLTÓ-like scenarios. It would be of interest to explore such moduli stabilization scenarios further and, in particular, to continue the search for susy-breaking vacua.

\section{Conclusions}\label{conclusions_sec}

In this paper we have developed a practical mechanism for
perturbatively stabilizing the complex structure moduli of a
Calabi-Yau threefold in the context of a heterotic
compactification. The remarkable feature of this mechanism is that the
moduli are stabilized while in no way changing the K\"ahler or
topological structure of the manifold. That is, we stabilize moduli
while staying within the class of Calabi-Yau manifolds. Unlike other
mechanisms \cite{Lukas:1996zq}-\cite{Braun:2006th} for stabilizing
moduli -- such as introducing topologically non-trivial flux -- which
dramatically change the properties of the base manifold, this approach
allows us to keep the toolkit of algebraic (K\"ahler) geometry. As a
result, we can hope to combine this mechanism with known approaches to
constructing realistic visible sectors
\cite{burt_a}-\cite{Anderson:2011ns}, rather than being forced to
start over with the difficult task of computing spectra (cohomology)
and designing phenomenologically relevant gauge bundles on new,
non-K\"ahler spaces.

\section*{Acknowledgments}
L.A. would like to thank T. Pantev for helpful conversations. L.A. and B.A.O. are supported in part by the DOE under contract number No. DE-AC02-76-ER-03071 and the NSF under RTG DMS-06366064 and NSF-1001296. A. L. is supported by the EC 6th Framework Programme MRTN-CT-2004-503369 and by the EPSRC network grant EP/l02784X/1. J.G. acknowledges support by the NSF-Microsoft grant NSF/CCF-1048082 and would like to thank the University of Pennsylvania for hospitality while part of this research was completed.

\appendix

\section{Deformations of the Total Space of a Bundle}\label{atiyah_total_space}

In this section, we give a brief review of the derivation of the Atiyah sequence by describing the deformations of the total space of $\pi: V \to X$ as a complex manifold (see \cite{donaldson_def} for a more complete discussion).

To measure the simultaneous infinitesimal deformations of $V$ and $X$ we are interested in the deformations of the total space (fiber $+$ base) of $V \stackrel{\pi}{\to} X$. If the total space were a compact manifold, then this analysis would be an example of what we have already discussed. It is well-known that the first-order deformations of any compact, complex space $X$ are simply measured by $H^1(X,TX)$ \cite{kodaira, kobayashi}. However, since the fiber directions of $V$ are non-compact, one must first reduce the problem to that of a compact space by considering separately the deformations of the line bundle $\Lambda^{top} V$ and the projective bundle ${\cal P}=\mathbb{P}(V) \stackrel{r}{\to} X$. From this decomposition, there arises the short exact sequence
\beq\label{upstairs_seq}
0 \to T_{{\cal P}|X} \to T{\cal P} \to r^{*}(TX) \to 0
\eeq
of bundles over ${\cal P}$, where $ T_{{\cal P}|X}$ describes the vertical (fiber) directions. The Leray spectral sequence associated with the map $r$ gives us \cite{donaldson_def}
\beq
H^i(P, r^*(TX))=H^i(X, TX)~\text{and}~~ H^i(P, T_{{\cal P}|X})=H^i(X, \textnormal{End}(V)) \ ,
\eeq
where $\textnormal{End}$ denotes the trace-free endomorphisms. In terms of these results, $H^1(X, T{\cal P})$ can be described by the long-exact sequence in cohomology
\beq
0 \to H^1(X, \textnormal{End}(V)) \to H^1(P, T{\cal P}) \to H^1(X, TX)  \stackrel{\alpha}{\to} H^2(X, \textnormal{End}(V)) \to \ldots \ .
\eeq
This is the sequence \eref{atiyah_coh} presented in Section \ref{full_atiyah}. It gives $H^1(P, T{\cal P})$, the first-order deformations of the pair $(X, V)$, in terms of the deformations of $V$ (with $X$ fixed) and the deformations of $X$ (regardless of $V$). 

Instead of using the Leray sequences to determine $H^1(P, T{\cal P})$, one could have considered directly the restriction of \eref{upstairs_seq} to the base $X$. Performing this restriction, we get a short exact sequence over $X$ given by 
\beq
0 \to  \textnormal{End}(V) \to {\cal Q} \to TX \to 0~~.
\eeq
This is the ``Atiyah sequence'' of \eref{atiyah_seq}. By definition, we have defined ${\cal Q}$ so that 
\beq
H^1(X, {\cal Q})=H^1(P, T{\cal P}) \ ,
\eeq
as described above. The coboundary map $\alpha=[F^{1,1}] \in H^1(X, V \otimes V^{\vee})$ is the Atiyah class \cite{atiyah,donaldson_def}.

\section{Moduli of Bundles Defined By Extension}\label{ext_moduli}
In this Appendix, we give a brief discussion of the bundle moduli associated with the $SU(2)$ bundle presented in Section \ref{main_eg}. Let $V$ defined by 
\beq\label{seq_again}
0 \to \cL \to V \to \cL^{\vee} \to 0
\eeq
be a holomorphic vector bundle with structure group $SU(2)$ defined over $X$, a Calabi-Yau threefold. The bundle $V$ is specified by the extension class $\phi \in H^1(X, \cL^2)$. For $V$ to be an example of the type discussed in Section \ref{main_eg}, the cohomology $H^1(X, \cL^2)$ containing the extension class must be able to  ``jump" in an index preserving manner with $H^2(X, \cL^2) \simeq H^1(X, {\cL^{\vee}}^2)$. Without loss of generality, we will assume that at our initial starting point in complex structure moduli space $H^1(X, \cL^2)$ and $H^1(X, {\cL^{\vee}}^2)$ are both non-vanishing. Furthermore, by construction, we choose the line bundle $\cL$ and the K\"ahler moduli so that $\mu(\cL) < 0$. For such a line bundle and region of moduli space, $V$ is slope-stable. Consistent with stability, one must have $H^0(X,\cL^2)=H^0(X, {\cL^{\vee}}^{2})=0$. 

To determine the local moduli space of $V$, we must compute the cohomology $H^1(X, \textnormal{End}(V))=H^1(X, V \otimes V^{\vee})$. This can be done using the array
\beq\ba{ccccccccc}\label{display_app}
&&0&&0&&0&& \\
&&\downarrow&&\downarrow&&\downarrow&& \\
0&\to& {\cal L}^{\otimes 2} &\to& V\otimes {\cal L} &\to& \cO_{X}&\to&0  \\
&& \downarrow &&\downarrow&&\downarrow&&\\
0&\to& {\cal L}\otimes V^{\vee}&\to& V\otimes V^{\vee} &\to& {\cal L}^{\vee}\otimes V^{\vee}&\to&0 \\
&&\downarrow&&\downarrow&&\downarrow&& \\
0&\to&\cO_X &\to&  {\cal L}^{\vee}\otimes V & \to & {{\cal L}^{\vee}}^{\otimes 2} &\to&0 \\
&&\downarrow&&\downarrow&&\downarrow&& \\ 
&&0&&0&&0&& \\ 
\ea \eeq 
 which follows directly from \eref{seq_again}.
To begin, consider the three short exact sequences (the first and last column and the middle row of the array in \eref{display_app})
\begin{align}
0 \to \cL \otimes V^{\vee} \to  & V \otimes V^{\vee} \to \cL^{\vee} \otimes V^{\vee} \to 0\label{modu1} \\
0 \to \cL^2 \to & \cL \otimes V^{\vee} \to \cO_{X} \to 0\label{modu2} \\
0 \to \cO_{X} \to & \cL^{\vee} \otimes V^{\vee} \to {\cL^{\vee}}^{2} \to 0\label{modu3} 
\end{align}

From these sequences, take the associated long exact sequences in cohomology, beginning with \eref{modu2}. These are
\begin{align}
&0 \to H^0(X, \cL \otimes V^{\vee}) \to H^0(X, \cO_{X}) \stackrel{\phi}{\rightarrow} H^1(X,\cL^2) \to H^1(X, \cL \otimes V^{\vee}) \to 0 \label{ahem1} \\
& H^2(X, \cL \otimes V^{\vee})\simeq  H^2(X, \cL^2) \\
&H^3(X, \cL \otimes V^{\vee}) \simeq H^3(X, \cO_{X}) \simeq \mathbb{C} \label{ahem4}
\end{align}
Since $\phi \in H^1(X, \cL^2)$ is a non-trivial extension class, the induced coboundary map $\phi$ in \eref{ahem1} is injective and $H^0(X, \cL \otimes V^{\vee})=0$. With this observation, \eref{ahem1} reduces to 
\beq
H^1(X, \cL \otimes V^{\vee})=\frac{H^1(X,L^2)}{\mathbb{C}} \ .
\eeq
Next, from \eref{modu3}, we note that 
\begin{align}
&\mathbb{C} \simeq H^0(X,\cO_X) \simeq H^0(X, \cL^{\vee} \otimes V^{\vee})\label{last_ahem1} \\
&H^1(X, \cL^{\vee} \otimes V^{\vee}) \simeq H^1(X,{\cL^{\vee}}^2) \\
&0 \to H^2(X, \cL^{\vee} \otimes V^{\vee}) \to H^2(X,{\cL^{\vee}}^2) \stackrel{\phi}{\rightarrow} H^3(X, \cO_X) \to H^3(X, \cL^{\vee} \otimes V^{\vee}) \to 0 \label{last_ahem4}
\end{align}
By a similar argument to that above, one can verify that $\phi$ in \eref{last_ahem4} is surjective and $H^3(X, \cL^{\vee} \otimes V^{\vee})=0$. Thus, we have
\beq
H^2(X, \cL^{\vee} \otimes V^{\vee}) = \textnormal{Ker} \left( \phi: H^2(X,{\cL^{\vee}}^2) \to H^3(X, \cO_X) \right)
\eeq
with dimension $h^2(X, \cL^{\vee} \otimes V^{\vee})=h^2(X,{\cL^{\vee}}^{2})-1$.

We can now substitute this information into the cohomology sequence for \eref{modu1} to determine the form of $H^1(X, V \otimes V^{\vee})$. We find
\begin{align}
&0\to H^0(X, V\otimes V^{\vee}) \to H^0(X,  \cL^{\vee} \otimes V^{\vee})\label{app1}   \\
&\stackrel{\delta}{\hookrightarrow} H^1(X, \cL\otimes V^{\vee}) \to H^1(X, V\otimes V^{\vee}) \to H^1(X, \cL^{\vee} \otimes V^{\vee})   \\
& \stackrel{\delta}{\hookrightarrow}  H^2(X, \cL\otimes V^{\vee}) \to H^2(X, V\otimes V^{\vee}) \to H^2(X, \cL^{\vee} \otimes V^{\vee}) \\
&\stackrel{\delta}{\hookrightarrow}  H^2(X, \cL\otimes V^{\vee}) \to H^3(X, V\otimes V^{\vee}) \to 0 \ ,\label{app4}
\end{align}
where the coboundary map, $\delta$, is an element of $H^1(X, \cL^2 \times V \otimes V^{\vee})$. Since $V$ is stable, $H^0(X, V\otimes V^{\vee})= \mathbb{C}$ and the filtration of $V$ by $\cL$ in \eref{seq_again} induces an identification of $\delta$ with the extension class, $\phi \in H^1(X, \cL^2)$. That is,
\beq
\delta =\phi \in H^1(X, \cL^2) \simeq H^0(X,  V \otimes V^{\vee}) \times H^1(X, \cL^2) \subset H^1(X, \cL^2 \times V \otimes V^{\vee}) \ .
\eeq
Using this coboundary map, and the results of \eref{ahem1}-\eref{ahem4} and \eref{last_ahem1}-\eref{last_ahem4}, the sequences in \eref{app1}-\eref{app4} reduce to 
\begin{align}
&H^0(X, V\otimes V^{\vee}) \simeq H^3(X, V\otimes V^{\vee}) \simeq \mathbb{C} \\
& H^1(X, V \otimes V^{\vee}) \simeq \frac{H^1(X, \cL^2)}{\mathbb{C}} \oplus H^1(X, {\cL^{\vee}}^2) \\
& H^2(X, V \otimes V^{\vee}) \simeq H^2(X, \cL^2) \oplus \textnormal{Ker}\left( \phi: H^2(X,{\cL^{\vee}}^2) \to \mathbb{C} \right)
\end{align}
We have, at last, computed the general dimension 
\beq
h^1(X, V\otimes V^{\vee})=h^2(X, V\otimes V^{\vee})=h^1(X, \cL^2) +h^1(X, {\cL^{\vee}}^2)-1
\eeq
of the vector bundle moduli space associated with $V$ in \eref{seq_again}.

\section{Scale Independence in Physical Descriptions of Complex
  Structure Moduli Space}\label{hom_scaling}

It is well known that complex structure moduli space can be described
in terms of a set of homogeneous coordinates ${\cal Z}^A$, with two
points being identified if they are related by complex rescaling
${\cal Z}^A \sim \Delta \: {\cal Z}^A$. Since two points in the space
of ${\cal Z}^A$ values are the same point in complex structure space
if related by such a scaling, $\Delta$ must completely drop out of
any physical description of the moduli space. This is often achieved
by working in so-called affine variables, but this is in fact
unnecessary.

Let us start by considering the K\"ahler potential on complex
structure moduli space \cite{Castellani:1990tp,Castellani:1990zd,Strominger:1990pd}
\begin{eqnarray} \label{mrK}
K_{\textnormal{CS}} = - \textnormal{ln} (i (\bar{{\cal Z}}^{A} {\cal G}_A - Z^A \bar{{\cal G}}_A)) \;.
\end{eqnarray}
Here we take the redundant set of fields ${\cal Z}^A$ to be the lowest
components of the relevant ${N}=1$ chiral superfields. The
holomorphic function ${\cal G}$ is called the ${N}=2$
prepotential and determines the metric on moduli space completely. 
A lower subscript denotes a derivative with respect to ${\cal Z}^A$. 
The prepotenial has scaling dimension $2$, that is, ${\cal G}(\Delta {\cal Z})
\sim \Delta^2 {\cal G}({\cal Z})$. Under such a scaling, the Kahler potential~\eqref{mrK}
changes by a Kahler transformation
\begin{eqnarray} \label{kscale}
  K_{\textnormal{CS}} \rightarrow K_{\textnormal{CS}} -
  \textnormal{ln}(\Delta) - \textnormal{ln}( \bar{\Delta}) \; ,
\end{eqnarray}
The supergravity action can be written entirely in terms of the function $G=K+\ln |W|^2$ and its derivatives. 
This function $G$ needs to be invariant under the above scaling in order for the supergravity action to be and, as a result, the superpotential has scaling dimension one and must change as
\begin{equation}  
 W\rightarrow \Delta W\; .
\end{equation} 
In particular, this means every superpotential term must depends on the homogeneous complex structure coordinates ${\cal Z}^A$. 

\section{Computational Details}\label{comp_detailsec}

In this Appendix, we briefly outline the details of the calculation which determined the number of moduli stabilized in Subsection \ref{equiv_downstairs}.  In the example presented in Section \ref{our_eg},  the holomorphy of the bundle
\beq\label{cool_bundle}
0 \to \cO(-3,3) \to V \to \cO(3,-3) \to 0
\eeq
was studied over the quotient manifold $X/(\mathbb{Z}_{3}\times \mathbb{Z}_3)$ where $X$ was defined in \eref{eg1cy3} by the vanishing of a bi-degree $(3,3)$ hypersurface in $\mathbb{P}^2\times \mathbb{P}^2$. As in Subsection \ref{equiv_downstairs}, we will label the coordinates of the ambient projective spaces via, $\{x_i, y_i\}$ where $i=0,1,2$ and the $\mathbb{Z}_3 \times \mathbb{Z}_3$ symmetry acts as on these homogeneous coordinates in \eref{z3z3}.

As in \eref{bicubic_poly}, the most general defining polynomial invariant under the discrete automorphism is
\begin{eqnarray}\label{bicubic_poly_app}
p_{(3,3)}&=&A_{1}^{k,\pm} \sum_{j} x_{j}^{2}x_{j\pm 1}y^{2}_{j+k}y_{j+k\pm1}+A_{2}^{k}\sum_{j}x^{3}_{j}y^{3}_{j+k}+A_{3}x_1x_2x_3\sum_{j}y^{3}_j\nonumber\\
&&+A_{4}y_1y_2y_3\sum_{j}x^{3}_{j}+A_5x_1x_2x_3y_1y_2y_3 \ ,
\end{eqnarray}
where $j,k=0,1,2$ and there are a total of $12$ free coefficients, denoted by $A$ with various indices. Subtracting the freedom of an overall scaling in the $p=0$ equation leaves $h^{2,1}(X/(\mathbb{Z}_3\times \mathbb{Z}_3))=11$.

In this Appendix, we explicitly demonstrate the fluctuation, or ``jumping extension calculation" for the bundle $V$ in \eref{bicubic_poly_app} on $X/(\mathbb{Z}_{3}\times \mathbb{Z}_3)$. This will be similar in structure to that described in \eref{fluct_again} for  the covering space, $X$ given in \eref{eg1cy3}. In particular, we must determine where in the complex structure moduli space of $X/(\mathbb{Z}_{3}\times \mathbb{Z}_3)$, the cohomology
\beq
H^2(X/(\mathbb{Z}_{3}\times \mathbb{Z}_3), \cL^2)=H^2(X/(\mathbb{Z}_{3}\times \mathbb{Z}_3), \cO(-6,6))
\eeq
is non-vanishing? This requires that we solve the fluctuation equation
\beq\label{fluct_baby}
(p_0 + \delta p)(k_0 + \delta k)=0
\eeq
where $p_0$ is an initial {\it invariant} defining polynomial of the form \eref{bicubic_poly_app} and $k_0$ is a starting element of $H^2(X, \cN^{\vee}\otimes L^2)=H^2(X, \cO(-9,3))$ which is invariant under the symmetry action given in \eref{z3z3}. To proceed a starting point $p_0,k_0$ must be chosen, the invariant fluctuations $\delta p, \delta k$ defined, and an explicit polynomial description of all of the above given. Before we can clearly describe this calculation then, we must first provide such a polynomial (algebraic) description of the cohomology. Thus, we will briefly detour from our goal here to first describe the relevant cohomology that we will need to analyze \eref{fluct_baby}

In this work, we will employ the Bott-Borel-Weil Formalism
\cite{Hubsch:1992nu}, for describing the cohomology $H^{\star}(X,
\cL)$ for any line bundle $\cL$. Once this description of the covering
space cohomology is obtained, we can provide a polynomial description
of $H^{\star}(X/(\mathbb{Z}_{3}\times \mathbb{Z}_3), \cL)$ by simply
finding the elements of this cohomology that are invariant under
\eref{z3z3}. It is beyond the scope of this work to provide a thorough
review of this formalism and here we merely direct the reader to more
complete discussions provided in
\cite{Hubsch:1992nu,Anderson:2008ex}. In the following paragraphs we
provide a brief summary of the key ideas.

The polynomial descriptions we require begin with observations about
line bundle cohomology over a single projective space. For any line
bundle defined over a single projective space, $\mathbb{P}^n$, the
only possibly non-vanishing cohomology groups are $H^0(\mathbb{P}^n,
\cL)$ and $H^n(\mathbb{P}^n, \cL)$ and only one of these can be
non-vanishing for a given line bundle.  If the homogeneous coordinates
of projective space are denoted $z_i$, $i=0,\ldots n-1$, then, for an
ample line bundle $\cO(k)$, $k\geq 0$, on $\mathbb{P}^n$, the
polynomial representation is nothing more than the global sections
\beq\label{h0_poly} H^0(\mathbb{P}^{n}, \cO(k)) \sim \{ z_{0}^{k},
z_0^{k-1}z_1, \ldots, z_{n-1}^{k} \} ~.  \eeq That is the set of all
homogeneous degree $k$ polynomials over $\mathbb{P}^n$. Likewise, for a
line bundle of the form $\cO(-k)$ with $k>n$, we can find a polynomial
description of $H^n(\mathbb{P}^{n},\cO(-k))$. In this case, however,
the description is in terms of ``inverse" polynomials\footnote{See \cite{Blumenhagen:2010pv} for related ideas.} of degree
$-(k-n)$ of the form \beq\label{htop_poly} H^n(\mathbb{P}^n, \cO(-k))
\sim \{\frac{1}{z_{0}^{k-n}}, \frac{1}{z_0^{k-n-1}z_1},
\ldots, \frac{1}{ z_{n-1}^{k-n}} \} ~.  \eeq

In this description, multiplication of regular and inverse polynomials takes the delta-function form that
\beq\label{trim_rule}
z_i^{k} \cdot \frac{1}{z_j^{k}} = \begin{cases}
1 & \text{if $i=j$} \\ 
0 & \text{if $i\neq j$}
\end{cases}~~.
\eeq
This concisely encodes the Serre duality mapping 
\beq
H^0(\mathbb{P}^n,\cO(k)) \times H^n(\mathbb{P}^{n}, \cO(-k-n-1)) \to H^n(\mathbb{P}^n, \cO(-n-1))=\mathbb{C}
\eeq
With this consistent polynomial representation in hand we can extend it to the case of interest -- multi-projective spaces -- via the Kunneth formula \cite{AG} which states that for any sheaf, $U$, defined on a direct product space $X \times Y$, the cohomology decomposes as
\beq
H^{k}(X\times Y, U) = \bigoplus_{k=i+j} H^i(X, U|_{X}) \otimes H^j(Y, U|_{Y})
\eeq

We are ready to represent the  cohomology relevant to
\eref{fluct_baby}. We begin with the cohomology on $X$ defined by
$p_{(3,3)}=0$. The Koszul sequence for $\cO(-6,6)$ gives us \beq 0 \to
H^1(X, \cO(-6,6)) \to H^2(\mathbb{P}^2\times \mathbb{P}^2, \cO(-9,3))
\stackrel{p_0}{\longrightarrow} H^2(\mathbb{P}^2\times \mathbb{P}^2,
\cO(-6,6)) \to H^2(X, \cO(-6,6)) \to 0 \eeq where by \eref{h0_poly},
\eref{htop_poly} and \eref{trim_rule},
\begin{align}
 H^2(\mathbb{P}^2\times \mathbb{P}^2, \cO(-9,3)) & =\text{space of homogeneous polynomials of bi-degree}~(-6,3) \\
 & =\{ \frac{y_{0}^{3}}{x_{0}^{6}},  \frac{y_{0}^{2}y_1}{x_{0}^{6}},\ldots  \frac{y_{0}^{3}}{x_{0}^{5}x_1}, \ldots \frac{y_{2}^{3}}{x_{2}^{6}} \} \\
H^2(\mathbb{P}^2\times \mathbb{P}^2, \cO(-6,6)) &= \text{space of homogeneous polynomials of bi-degree}~(-3,6) \\
 & =\{ \frac{y_{0}^{6}}{x_{0}^{3}},  \frac{y_{0}^{5}y_1}{x_{0}^{3}},\ldots  \frac{y_{0}^{6}}{x_{0}^{2}x_1}, \ldots \frac{y_{2}^{6}}{x_{2}^{3}} \}
\end{align}
and $p_0$ is the degree $(3,3)$ defining polynomial. Note that \eref{trim_rule} implies for example that the multiplication $ \frac{y_{2}^{3}}{x_{2}^{6}} \times x_0^3y_0^3$ maps to zero in $H^2(\mathbb{P}^2\times \mathbb{P}^2, \cO(-6,6)) $ since the monomials in the numerator and denominator are of the wrong type to cancel out and map into an appropriate element of the target.

We choose as the initial point in complex structure moduli space the defining polynomial
\bea\label{rdpoly}
p_0=  x_1^3 y_0^3 - x_2^3 y_{0}^{3} - x_0^3y_1^3 + x_2^3 y_1^3 + 
x_0^3 y_2^3 - x_1^3 y_2^3
\eea
corresponding to the coefficient choice $A_2^{2}=1$, $A_2^{1}=-1$ and all others vanishing. 

The invariant basis of $H^2(\mathbb{P}^2 \times \mathbb{P}^2,
\cO(-9,3))$ contains $33$ elements. We choose a random element of this
space as our starting point $k_0$. Next, to determine for which
fluctuations of complex structure $H^1(X, \cO(-6,6))$ can remain
non-vanishing, we must form the ideal given by \eref{fluct_baby} and
eliminate the $\delta k$ degrees of freedom in order to find the
allowable fluctuations $\delta p$. This can be achieved using a
Groebner basis computation in any appropriate elimination ordering
\cite{Gray:2008zs}.

We find at the given starting point in \eref{rdpoly} there are only
two degrees of freedom in $\delta p$ that will solve
\eref{fluct_baby}. This fluctuation of the defining polynomial is given by 
\bea\label{give_the_free_param}
\delta p=\delta A_2^{2} x_0^3 y_0^3 + (1 - \delta A_{2}^{1}- \delta A_2^{2}) x_1^3 y_0^3 + (-1 + \delta A_{2}^{1}) x_2^3 y_0^3 + (-1 + \delta A_{2}^{1}) x_0^3 y_0^3 +
\delta A_2^{2}x_1^3 y_1^3  \\ \nonumber
+ (1 -\delta A_{2}^{1} - \delta A_2^{2}) x_2^3 y_1^3 + 
(1 - \delta A_{2}^{1}- \delta A_2^{2}) x_0^3 y_2^3 + (-1 + \delta A_{2}^{1}) x_1^3 y_2^3 + 
\delta A_2^{2}x_2^3 y_2^3
\eea
One of the two remaining degrees of freedom is the overall scale, which drops out and is not
a complex structure modulus. As a result, this local calculation
implies that the presence of the holomorphic bundle $V$ in \eref{cool_bundle} stabilizes $10$ out of the $11$ moduli. 


\end{document}